\documentclass{aa}

\usepackage{graphicx}
\usepackage{amsmath}
\usepackage{xcolor}
\usepackage{url}

\usepackage{txfonts}
\usepackage{hyperref}

\def \cesamxx{Cesam2k20\xspace}
\def \eos{EoS}

\def \Tunit{~\mbox{K}}
\def \omegaunit{\mbox{rad}\,\mbox{s}^{-1}}
\newcommand{\Ross}{\operatorname{{R\kern-.04em o}}}
\def \xX{\mathrm{X}}

\bibpunct{(}{)}{;}{a}{}{,}
\usepackage[switch]{lineno}
\begin{document}

   \title{Water gas discs in exo-asteroid belts}

   \subtitle{}

   \author{Paul Huet\inst{1}\thanks{E-mail: paul.huet@obspm.fr} \and Quentin Kral\inst{1} \and Louis Manchon\inst{1}} 
   
   \institute{LIRA, Observatoire de Paris, Université PSL, Sorbonne Université, Université Paris Cité, CY Cergy Paris Université, CNRS, 92190 Meudon, France
   }

   \date{}

  \abstract
{Observations of tens of secondary CO gas discs associated with cold exo-Kuiper belts together with other arguments have led \cite{kral_impact-free_2024} to propose that water ice could also sublimate in exo-asteroid belts, suggesting a new pathway for the delivery of water to terrestrial planets, including Earth.}
{We aim to model such water vapour discs and to characterise their physical properties across a range of extrasolar systems with different host stars. We further investigate the implications for the accretion of this water by potential planets located in the inner regions of these systems.}
{We adapt and extend the gas production and evolution model of \cite{kral_impact-free_2024} to follow the outgassing, photodissociation, and viscous evolution of water vapour discs. We perform a suite of simulations exploring the parameter space, focusing on the stellar mass, the mass of the parent belt, and its orbital location. We additionally include an inner planet to estimate the mass of water accreted as a function of disc properties and system architecture.}
{We find that systems hosting more massive stars ($M_\star \gtrsim 1\,M_\odot$) produce water vapour very efficiently, sublimating nearly all of the ice initially present in the belt. In most cases, the bulk of the gas mass is generated early, when the stellar luminosity is highest. Long-term shielding of water vapour against photodissociation occurs only in the most massive belts. The amount of water accreted by inner planets can approach the initial ice mass of the belt, leading to planets with water inventories comparable to or exceeding those of Earth, potentially creating ocean planets.}
{We find that water outgassing occurs early after the protoplanetary disc dissipates in systems containing exo-asteroid belts. ALMA, JWST and ELT are capable of detecting this water vapour for several tens of millions of years, even in relatively low-mass water discs containing down to a few $10^{-9}$ M$_\oplus$ at 100 pc. Hence, if such water gas discs are present, they should be detectable with current facilities.
}
  {}
   \maketitle

\section{Introduction}
The supply of water to planets is one of the most important factors in determining a planet's habitability \citep{kasting_habitable_1993}. Recently, a new mechanism has been proposed for supplying water from the spread of a disc of water vapour produced in an exo-asteroid belt, without the need for impacts \citep{kral_impact-free_2024}. This mechanism has been tested for the solar system, but as it is a priori universal, it could also apply to extrasolar systems, which we will test further in this paper. 

Outgassing in belts around main-sequence stars has been observed in tens of systems \citep[e.g.][]{moor_molecular_2017, matra_detection_2017, moor_new_2019, matra_resolved_2025}. Indeed, degassing of cold extrasolar belts resembling the Kuiper belt is now well studied, both observationally \citep[see e.g., ][]{matra_detection_2017, matra_resolved_2025, marino_population_2020, cataldi_surprisingly_2020} and theoretically \citep[see e.g., ][]{kral_self-consistent_2016,kral_predictions_2017,kral_imaging_2019,bonsor_secondary_2023, cataldi_primordial_2023}.
For now, water has never been observed outgassing in those types of belts but thanks to the high resolution and great sensitivity of ALMA, CO and carbon gas have been detected in many extrasolar Kuiper belts {\citep[see e.g., ][]{moor_molecular_2017, moor_new_2019,matra_resolved_2025, lu_jwstnirspec_2026}.

These belts or debris discs, which are older and lighter than protoplanetary discs, are the remnants of planetary formation \citep{hughes_radial_2017, armitage_astrophysics_2020}. The primordial gas in protoplanetary discs, composed of a mixture of hydrogen and helium, is expected to dissipate within a few million years {\citep{ribas_disk_2014, ribas_protoplanetary_2015, polnitzky_precise_2026}. On the other hand, debris discs can last for Gyr, and though their solid mass content is lighter than in their protoplanetary disc counterparts, they are typically much heavier than that of the current Kuiper belt \citep{matra_resolved_2025}. These cold discs are located close to the CO ice line \citep{matra_empirical_2018, matra_resolved_2025}, and models suggest that the observed CO gas is secondary \citep{kral_predictions_2017, kral_imaging_2019}. The gas release mechanism is not yet fully understood \citep[e.g.][]{bonsor_secondary_2023}, but it could originate in collisions between solids, allowing the release of volatile ice contained within planetesimals, or by thermal heating from the central star  \citep{kral_predictions_2017, kral_molecular_2021, huet_late_2025}.

The presence of late gas in planetary systems for extended periods is expected to have a significant impact. Indeed, this gas could efficiently be accreted onto planets \citep{kral_formation_2020}. For example, the Solar System may have hosted a secondary gas disc as a consequence of volatile gas released in the originally more massive Kuiper belt. \citet{huet_late_2025} show that this potential late gas disc leads to significant accretion onto Uranus and Neptune, which may have significantly contributed to the high atmospheric C/H ratio observed ($\gtrsim 60$ times solar) for both ice giants.

In the same vein, if water gas was released in the young and massive asteroid belt, it could explain the total quantity of water (on the surface and mantle) observed on Earth and other terrestrial planets of the Solar System \citep{kral_impact-free_2024}. It is well possible that exo-asteroid belts behave like their colder exo-Kuiper belt counterparts and outgas their volatiles. 
Those warmer belts, close to the water ice-line, are observed in many systems \citep{morales_common_2011, chen_spitzer_2014,ballering_what_2017}, and it is expected that they would rather outgas water than CO \citep{kral_impact-free_2024}.

Direct observations of water from the ground appear more challenging than those of CO. Indeed, the presence of water in the atmosphere increases the integration time for radioastronomical observations \citep{facchini_resolved_2024}. Water gas discs are also warmer than CO gas discs, which makes them harder to observe at millimetre wavelengths. However, gaseous water has already been observed with ALMA in protoplanetary discs \citep{facchini_resolved_2024}. Water ice has also recently been observed in a debris disc by JWST \citep{xie_water_2025}, which was expected for a long time.

The presence of such secondary water gas discs could significantly alter our understanding of the evolution of extrasolar systems. The accretion of water onto inner planets, sometimes close to the habitable zone, could significantly affect their environment \citep{kral_impact-free_2024}. For instance, a completely dry terrestrial planet with the right temperature could acquire a massive ocean \citep{kral_formation_2020}

The objective of this paper is to build a model that can estimate which planetary systems can host secondary water gas discs, as well as estimate their masses and observability for different stellar hosts. We will also investigate the amount of water potentially accreted by an inner planet with the mass of Earth in these systems. The final goals are to use these results to guide us towards the first detection of these water gas discs and explore the universality of the mechanism, which could have strong implications for the development of life in extrasolar systems.

\section{Method}

Our numerical model is an improved version of the code described in \cite{huet_late_2025}, adapted for water sublimation based on the work of \cite{kral_impact-free_2024}. The numerical model, \texttt{Diffenix}\footnote{\url{https://huetpaul.github.io/Diffenix}} is publicly available online with its full documentation.

\subsection{Properties of the exo-asteroid belts}\label{sub_seq:ast_belt}

The exo-asteroid belts modelled in this study extend from \(a_\text{belt} - \Delta a / 2\) to \(a_\text{belt} + \Delta a / 2\), where \(a_\text{belt}\) and \(\Delta a\) are the mean semi-major axis and the width of the belt, respectively. \(\Delta a\) is chosen such that \(\frac{\Delta a}{a_\text{belt}} = 0.5\) to match the asteroid belt of the solar system and the belt modelled in \cite{kral_impact-free_2024}. It is also in line with observations of debris discs \citep{matra_resolved_2025}.

The radial distribution is defined as in \cite{kral_impact-free_2024} using the Minimum Mass Solar Nebula (MMSN) model, with the density varying as \(r^{-3/2}\). The size distribution is assumed to be similar to that currently observed in the asteroid belt \citep{bottke_fossilized_2005}. Three power laws are used: between \(s_{\text{min}} = 1 \, \text{m}\) and \(s_{\text{med}} = 20 \, \text{km}\), the slope is \(q_{\text{low}} = 3.6\); between \(s_{\text{med}}\) and \(s_{\text{big}}\)  the slope is \(q_{\text{med}} = 1.2\), and between \(s_{\text{big}}\) and \(s_{\text{max}} = 1000 \, \text{km}\), the slope is \(q_{\text{high}} = 4.5\). The entire distribution $n(a, s)$, with $a$ the distance to the central star and $s$ the asteroid size, is then normalized to obtain a total mass $M_{\rm belt}$.

Since the luminosity of the star varies significantly depending on its mass, we did not choose an identical location for all the modelled belts. Instead, we used the protoplanetary disc ice-line at the time of its dissipation to calculate the location of the belts. In practice, we have chosen belts located at 0.7, 1.0, and 1.3 times the position of the ice-line (see Section~\ref{sub_sec:simulation_set}). This choice guarantees a similar insolation for all belts to be able to compare them.

To calculate the location of the ice-line in the protoplanetary disc, we use the midplane temperature as a function of the distance \(r\) from the central star, as given by Equation 15 of \cite{kral_impact-free_2024}:
\begin{equation}
    T^4 = T^4_{\text{irr}} + T^4_{\text{acc}}
\end{equation}
\noindent
where \(T^4_{\text{irr}}\) is the contribution from radiative heating \citep{chiang_spectral_1997, sasselov_snow_2000}:
\[
T^4_{\text{irr}} = \left(0.1 \frac{R_*}{r} + \frac{2}{7} \left(\frac{\text{k}_\text{B}^4 R_*^2 L_*}{8 \pi^2 \sigma \mu_* \mathcal{G}^4 \text{M}_*^4}\right)^{1/7} \left(\frac{r}{R_*}\right)^{2/7} \right) \left(\frac{R_*}{r}\right)^2 \frac{L_*}{8 \pi^2 \sigma R_*^2},
\]
with \(R_*\), \(\mu_*\), and \(L_*\) being the radius, mean molecular weight of the gas, and luminosity of the central star, respectively. As for \(\sigma\), \(\text{k}_\text{B}\), and \(\mathcal{G}\), they are the Stefan-Boltzmann, Boltzmann, and gravitational constant, respectively (see also \citealt{chiang_spectral_1997}). The term \(T^4_{\text{acc}}\) corresponds to the viscous heating associated to an accretion rate onto the central star \(\dot{M}_{\text{acc}}\) \citep{lecar_location_2006} :
\[
T^4_{\text{accr}} = \left[\frac{3}{4}\left(\tau_R + \frac{2}{3}\right)\right] \frac{3}{8 \pi \sigma} \frac{G \text{M}_* \dot{M}_{\text{accr}}}{r^3} \left(1 - \sqrt{\frac{R_*}{r}}\right),
\]
where \(\tau_R = \kappa_R \Sigma_\text{MMSN}\) is the Rosseland optical depth. We take \(\kappa_R \sim 1 \ \text{m}^2.\text{kg}^{-1}\) and \(\Sigma_\text{MMSN} = 1.7 \times 10^4 (r / (1 \ \text{au}))^{-3/2} \ \text{kg}.\text{m}^{-2}\) \citep{hayashi_structure_1981}.

Then, we find the location of the ice-line, \(a_{\text{iceline}}\) by solving the equation \(T(a_{\text{iceline}}) = 170 \, \text{K}\) for our different setups. We assume that in the last phase of the protoplanetary disc stage, $\dot{M}_{\text{accr}}=2 \ \times 10^{-9} \ \text{M}_\odot.\text{yr}^{-1}$, similar to \citet{kral_impact-free_2024}.

\subsection{Viscous diffusion} \label{sub_seq:viscous_diff}

The modelling of the dynamics of gas discs \citep[e.g.,][]{shakura_black_1973} over millions of years has progressively been enriched to model specific features, such as volatile ice sublimation and photodissociation, which are primordial ingredients for modelling debris discs \citep[e.g.][]{moor_new_2019, marino_population_2020}.

In our model, the disc is assumed to be axisymmetric and vertically isothermal. The typical vertical scale height provided by hydrostatic equilibrium is then $\text{H}_\text{z}=c_s/\Omega_\text{K}$ where $\Omega_\text{K}^2 = \mathcal{G}\text{M}_*/r^3$  is the local Keplerian angular speed and $c_s^2 = \text{k}_\text{B} T/(\mu \text{m}_\text{p})$ is the isothermal speed of sound, which depends on both temperature $T(r)$ and mean molecular mass, $\mu \text{m}_\text{p}$. 

The viscous parameter that sets the timescale of the gas dynamics can then be defined as $\alpha = \nu/(c_s \text{H}_\text{z})$ \citep{shakura_black_1973}, where $\nu$ is the gas viscosity, which is the sum of molecular and turbulent viscosities. Although there is not yet a fully self-consistent model to calculate $\alpha$ for debris discs, \cite{kral_magnetorotational_2016} and \cite{cui_dynamics_2024} have suggested that it could be relatively high (from $\sim 10^{-3}$ to $\sim 10^{-1}$ compared to values in protoplanetary discs. This is mainly due to the higher ionisation rates in debris discs, allowing for the magnetorotational instability to become very active \citep{kral_magnetorotational_2016}, and to some potential molecular viscosity that can appear at low gas densities \citep{cui_dynamics_2024}.
This viscosity induces a torque that can be introduced into the momentum conservation equation. Coupled with mass conservation, this gives the typical equation for the evolution of gas surface density $\Sigma(r, t)$ in cylindrical coordinates \citep{lynden-bell_evolution_1974, pringle_accretion_1981}

\begin{equation}
    \frac{\partial \Sigma}{\partial t} = \frac{3}{r} \frac{\partial}{\partial r} \left[\sqrt{r} \frac{\partial}{\partial r} \left[ \nu \Sigma \sqrt{r}\right]\right] + \dot{\Sigma},
    \label{eq:radial_diff}
\end{equation}

\noindent where the radial speed $v_r$ is given by
\begin{equation}
    \Sigma v_r = -\frac{3}{\sqrt{r}} \frac{\partial}{\partial r} \left[\nu \Sigma \sqrt{r}\right]. \label{eq:radial_speed}
\end{equation}

As is described in \citet[][]{kral_impact-free_2024} and \citet{huet_late_2025}, the source/sink term $\dot{\Sigma} = \dot{\Sigma}_\text{belt} +\dot{\Sigma}_\text{accr}$ contains all the physics specific to debris discs. The term \(\dot{\Sigma}_\text{belt}\) accounts for gas production by the belt. It is positive at the belt's location, where the gas is originally produced from ice sublimation and equal to zero everywhere else. On the other hand, the term \(\dot{\Sigma}_\text{accr}\) accounts for gas accretion by planets. Therefore, this term is negative at the planets' locations.

Similar to \citet[][]{kral_formation_2020, kral_impact-free_2024}, we assume the viscous parameter $\alpha$ to be a constant throughout disc. The temperature profile is given by $T(r, t) = T_0 \left(L_*(t)/L_\odot\right)^{1 / 4} \left(r/r_{\rm au}\right)^{\alpha_T}$ where $T_0 = 278$ K, $r_{\rm au} = 1 \ \mathrm{au}$, $L_*(t)$ is the central star's luminosity at time t and $L_\odot$ the current solar luminosity. The parameter $\alpha_T$ sets the steepness of the radial power law, usually assumed to be $-\frac{1}{2}$ \citep[e.g.][]{marino_population_2020, kral_impact-free_2024}.

With these assumptions, Eq.~\eqref{eq:radial_diff}  can be rewritten as a simpler diffusion equation that can be efficiently integrated numerically (see Appendix \ref{sub_seq:f_evolution}). To do so, we introduce $f=\Sigma r^{2+\alpha_T}$ and a new radial coordinate\footnote{One can then prove that the mass flux $\dot{m}(r)=2 \pi r \Sigma v_r$ is equal to $-4 \pi  \frac{3 \nu_0}{4 x_0^2} \frac{\partial f}{\partial x}$.} $x=\sqrt{r}$ so that

\begin{equation}
    \frac{\partial f}{\partial t} = D \frac{\partial^2 f}{\partial x^2} + \dot{f}
    \label{eq:diff}
\end{equation}

\noindent with $\dot{f}=x^{4 + 2 \alpha_T} \dot{\Sigma}$ and $D=3 \nu_0/(4 r_0^{3/2 + \alpha_T}) x^{1+2\alpha_t}$, where
$\nu_0~=~{\text{k}_\text{B} T(r_0) \ \alpha \ r_0^{3/2}}/({\mu \text{m}_\text{p} \sqrt{\mathcal{G}\text{M}_*}})$ is the viscosity at the arbitrary radius $r_0$, with $\mu \text{m}_\text{p}$ the mean molecular mass.
Assuming $\alpha_T = -0.5$, the diffusion coefficient $D$ becomes independent of the radial coordinate $x$ such that :
\begin{equation}
    D = \frac{3 \nu_0}{4 x_0^2}.
\end{equation}

The boundary conditions are the same as those used in \cite{huet_late_2025}, which were derived from \cite{marino_population_2020}. We use a power-law extrapolation to derive the outermost values and the minimum between a power-law extrapolation and a constant flux through the innermost two cells to derive the innermost values. 

In this paper, we model exo-asteroid belts and hence, assume that the gas produced within the belt is water. However, due to photodissociation caused by the central star and the interstellar radiation, this water can turn into neutral oxygen (O) and hydrogen (H). Estimating the fraction of water in the disc is important for determining whether it can be observed in its molecular form. To do so, we integrate the full set of equations for the evolution of $\Sigma_i$, the surface density of each species $i$, as described in \cite{huet_late_2025}. Here, $i$ is a shortcut for H$_2$O, H and O. The radial speed of a species $i$ is then the sum of the radial speeds calculated for the total surface density $v_r$ and the interspecies radial speed $v_r^i=-\nu \frac{\partial \omega_i}{\partial r}$, with $\omega_i = \frac{\Sigma_i}{\Sigma}$ \citep{charnoz_planetesimal_2019}.
Then each \(\Sigma_i\) satisfies:
 
\begin{equation}
    \frac{\partial \Sigma_i}{\partial t} = \frac{1}{r} \frac{\partial}{\partial r} \left[3 \omega_i\sqrt{r} \frac{\partial}{\partial r} \left[ \nu \Sigma \sqrt{r}\right] + r \nu \Sigma \frac{\partial \omega_i}{\partial r}\right] + \dot{\Sigma_i},
    \label{eq:sigma_i}
\end{equation}
\noindent with \(\dot{\Sigma}_i\) the sink/source term for the chemical species $i$.

Rewriting Eq. \eqref{eq:sigma_i} as a function of $f$, $\omega_i$ and $x$, yields to the typical advection-diffusion equation that follows
\begin{equation}
    \begin{split}      
    \frac{\partial \omega_i}{\partial t} &= \frac{D}{3} \left[\frac{\partial^2 \omega_i}{\partial x^2} 
    + 4 \frac{\partial \omega_i}{\partial x} \frac{\partial \ln{f}}{\partial x}\right] + \dot{\omega_i} \\
    &= \frac{\text{D}}{3} \frac{\partial^2 \omega_i}{\partial x^2} - \frac{2 x}{3} \text{v}_\text{r} \frac{\partial \omega_i}{\partial x} + \dot{\omega_i},    
    \end{split}
    \label{eq:omega_i}    
\end{equation}
\noindent where $\dot{\omega_i} = \frac{\dot{\Sigma_i} - \omega_i \dot{\Sigma}}{\Sigma}$. The term $\dot{\omega_i}$ is the sum of the contributions of two effects. The first one is \(\dot{\omega}_i^\text{belt}\) coming from water produced in the belt (see Section \ref{sub_sec:ice_sublimation}), and the other one \(\dot{\omega}_i^\text{photo}\) is due to photodissociation that decreases the amount of water over time (see Section \ref{sub_sec:photodissociation}). There is no contribution from gas accretion by the planet, as the efficiency of accretion does not depend on the chemical species considered (see section \ref{sub_sec:planetary_accr}).

To solve both Eqs. \eqref{eq:diff} and \eqref{eq:omega_i}, we follow the same procedure as in \cite{huet_late_2025} and discretize the equations spatially to integrate them in time using the \texttt{solve\_ivp} function of \texttt{SciPy} \citep[see ][]{virtanen_scipy_2020} with an \texttt{RK45} integration. Since Eq. \eqref{eq:diff} does not require $\omega_i$, but Eq. \eqref{eq:omega_i} does require f, we first integrate Eq. \eqref{eq:diff} and interpolate the result in time for the integration of Eq. \eqref{eq:omega_i}. With this method, those equations can be integrated numerically despite the difference between the characteristic timescales of photodissociation for different species and the timescales of gas production and diffusion. 

Estimates of the source/sink terms $\dot{f}$ and $\dot{\omega}_i$ will be further described in sections \ref{sub_sec:planetary_accr}, \ref{sub_sec:photodissociation}, and \ref{sub_sec:ice_sublimation}.

\subsection{Planetary accretion} \label{sub_sec:planetary_accr}
In this model, atmospheric accretion onto the planet is modelled via the $\dot{f}$ parameter in Eq. \eqref{eq:diff} using sink cells centred at the planet's semi-major axis $a_{\rm p}$ \citep[as in][]{kral_impact-free_2024, huet_late_2025}. The accretion rate may be slowed by two processes: thermal contraction of the atmosphere, which typically operates over large timescales \citep{piso_minimum_2014, lee_cool_2015}, and hydrodynamic processes, which operate in timescales much smaller than those involved in our one-dimensional viscous diffusion equation.

Quasi-hydrostatic contraction of the atmosphere was modelled by \cite{kral_formation_2020} for late gas in debris discs. The mass flux arriving at the planet is much smaller than that of protoplanetary discs and is insufficient to maintain a balance between the atmospheric heating by accretion and cooling by radiation. The accretion rate is therefore limited by the amount of matter falling onto the planet and thus by the mass flux crossing the planet's orbit as well as hydrodynamical effects. 
To model such complex hydrodynamics in our long-term one-dimensional viscous evolution equation, we parametrize the problem by assuming that the accretion rate is proportional to the mass flux crossing the planet's orbit, similar to \cite{kral_impact-free_2024, huet_late_2025} such that 
\begin{equation}
    \dot{M} = f_{\rm accr} \Phi_{\rm in},
\end{equation}
\noindent where $\Phi_{\rm in} = \min(1, \text{R}_\text{H} / \text{H}_\text{z}) \ 2 \pi a_{\rm p} \ \Sigma(a_{\rm p}, t) \ v_r(a_{\rm p}, t)$ is the flux arriving onto the planet located at a semi-major axis $a_{\rm p}$, and $f_{\rm accr}$ is the accreted fraction of that incoming flux.
$\text{R}_\text{H}$ and $\text{H}_\text{z}$ are the planet's Hill radius and the disc's vertical scale height, respectively. If the latter is greater than the maximum zone of influence of the planet's Hill radius, the flux available for accretion becomes smaller than the total mass flux crossing the planet's orbit as described by the factor $\min(1, \text{R}_\text{H} / \text{H}_\text{z})$ \citep{kral_impact-free_2024}. In this case, a fraction of the incoming gas, $1 - \text{R}_\text{H} / \text{H}_\text{z}$, passes over the planet's zone of influence and cannot be accreted.

Similar to \cite{kral_impact-free_2024}, based on 3D-hydrodynamical studies \citep{lubow_gas_2006, mordasini_characterization_2012, tanigawa_distribution_2012, ormel_hydrodynamics_2015, lambrechts_quasi-static_2019}, we set $f_{\rm accr} = 0.5$ for all simulations in this work.
Furthermore, the efficiency of accretion does not depend on the chemical element being accreted. Thus: $\dot{\Sigma}_i=\omega_i \dot{\Sigma}$, and $\dot{\omega}_i = 0$. 

\subsection{Photodissociation} \label{sub_sec:photodissociation}

Water gas can be photodissociated by UV photons from both the host star and the interstellar medium. To account for that, we have improved the model of \cite{kral_impact-free_2024} by adapting the method used by \cite{kral_predictions_2017, marino_population_2020, huet_late_2025}. We note that the main difference between secondary CO gas and H$_2$O is that the latter is only shielded by itself, while CO can be self-shielded or shielded by its daughter species, specifically neutral carbon \citep{kral_imaging_2019}.
The photodissociation of gaseous water is then modelled by
\begin{equation}
    \dot{\Sigma}_{\text{H}_2\text{O}} = - \frac{\Sigma_{\text{H}_2\text{O}}}{t_\text{ph}}
\end{equation}
\noindent with $t_\text{ph}$ the photodissociation timescale of water.
   The photodissociation timescale is hence the typical lifetime of a water molecule before being photodissociated.
   This timescale takes into account the shielding effect by the other molecules (see equation~\ref{eq:def_t_ph})
   As a result, H$_2$O is turned into neutral H and O, and we obtain
\begin{subequations}
    \begin{align}
        \dot{\Sigma}_{\rm O} &= \frac{\Sigma_{\text{H}_2\text{O}}}{t_\text{ph}} \frac{\mu_O}{\mu_{\text{H}_2\text{O}}}, \\
        \dot{\Sigma}_{\rm H} &= \frac{\Sigma_{\text{H}_2\text{O}}}{t_\text{ph}} \frac{2 \mu_H}{\mu_{\text{H}_2\text{O}}},
    \end{align}
\end{subequations}

\noindent where $\mu_{\text{H}_2\text{O}}$, $\mu_O$, and $\mu_H$ are the mean molecular weights of water, oxygen and hydrogen respectively.

The photodissociation conserves the total mass, hence we get the source term of Eq. \eqref{eq:radial_diff} due to photodissociation $\dot{\Sigma}_{\rm photo} = 0$ and 
\begin{subequations}
    \begin{align}
        \dot{\omega}_{\text{H}_2\text{O}}^{\rm photo} &= - \frac{\omega_{\text{H}_2\text{O}}}{t_\text{ph}}    \\
        \dot{\omega}_{\rm O}^{\rm photo} &= - \frac{\omega_{\text{H}_2\text{O}}}{t_\text{ph}} \frac{\mu_O}{\mu_{\text{H}_2\text{O}}}\\
        \dot{\omega}_{\rm H}^{\rm photo} &= \frac{\omega_{\text{H}_2\text{O}}}{t_\text{ph}}  \frac{2 \mu_H}{\mu_{\text{H}_2\text{O}}}
    \end{align}
\end{subequations}

\noindent where $\dot{\omega}_{\text{H}_2\text{O}}^{\rm photo}$, $\dot{\omega}_{\rm O}^{\rm photo}$ and $\dot{\omega}_{\rm H}^{\rm photo}$ are the source terms of Eq. \eqref{eq:omega_i} due to photodissociation.

$\dot{\omega}^{\text{photo}}$ is decomposed into two contributions, $\dot{\omega}^{\text{\rm photo}, *}$ and $\dot{\omega}^{\text{photo}, \text{ISRF}}$, due to photodissociation by the central star and the interstellar radiation field (ISRF), respectively. For each of these contributions, the photodissociation timescale is calculated differently, as the radiation intensity and the geometry of the problem vary.

The photodissociation timescale is chosen as follows \citep{kral_impact-free_2024}
\begin{equation}
    t_\text{ph}(r, t) = t_\text{ph}^{\rm f}(r, t) \exp{\left(\frac{N_{\text{H}_2\text{O}} \ \mu_{\text{H}_2\text{O}} \ \text{m}_\text{p}}{\Sigma_{\rm crit}}\right)}
    \label{eq:def_t_ph}
\end{equation}
\noindent where $t_\text{ph}^{\rm f}$ is the typical lifetime of a free water molecule, with no shielding effect from either stellar or interstellar radiations, and $N_{\text{H}_2\text{O}}$ is the number column density of water on the line of sight between the molecule and the impinging photon.
   The exponential term is due to the shielding effect by the other molecules. More precisely, when photodissociation operates, it decreases the flux of photons able to photodissociate molecules located further away. The sum of all of these absorptions is a decreasing UV flux following the Beer-Lambert law resulting in this exponential term in the equation~\ref{eq:def_t_ph}.
   Due to the geometry, a significant portion of the disc may then be optically thick to stellar UV photons but not to interstellar medium photons, which we account for in our model.
The critical surface density $\Sigma_{\rm crit}$, defined by \cite{kral_impact-free_2024} to be the surface density at which the optical depth $\tau_{\rm w}=N_{\text{H}_2\text{O}}\ \sigma_{\rm w}$ is equal to 1, is estimated from the water vapour absorption cross-section assumed to be a constant $\sigma_{\text{H}_2\text{O}} = 5 \times 10^{-22}$ m$^2$ in the UV \citep{bethell_formation_2009}. Then\footnote{Note that there is a typo in \cite{kral_impact-free_2024} where $\sigma_{\text{H}_2\text{O}}$ is correct, but there is a factor 10 difference in the final value of $\Sigma_{crit}$ they derive.} $\Sigma_{crit} = \frac{\mu_{\text{H}_2\text{O}} \,  \text{m}_\text{p}}{\sigma_{\text{H}_2\text{O}}}\simeq6 \times 10^5 \ \mathrm{kg.m^{-2}} \simeq 2.2 \times 10^{-7} \ \mathrm{M_\oplus.au^{-2}}$.

 \subsubsection{The central star's contribution}
To calculate the characteristic photodissociation timescale due to the central star at time \( t \) and distance \( r \), we first need to compute the characteristic photodissociation timescale for a single molecule, \(\text{t}_\text{ph}^{\text{f} \, *}(r, t)\).

If \(\sigma_{\rm w}(\lambda)\) is the effective cross-section for the photodissociation of water and \(\text{L}_\star(\lambda, t)\) is the stellar spectrum at time \( t \), then:

\begin{equation}
    \text{t}_\text{ph}^{\text{f} \, *}(r, t) = \left[ \int \frac{\text{L}_*(\lambda, t) \, \lambda}{\text{h}\text{c} \, 4 \pi r^2} \sigma_{\rm w}(\lambda) \, d\lambda \right]^{-1}
\end{equation}

\noindent Here, \({\text{L}_*(\lambda, t) \, \lambda}/({\text{h}\text{c} \, 4 \pi r^2})\) represents the radiation field of the central star at radius \( r \) and time \( t \) and h the Planck constant.

We assume  \(\sigma_{\rm w}(\lambda)\) is constant over the interval \(\Delta\lambda_\text{UV} = 91.2 - 200 \, \text{nm}\) (the integration domain), and we approximate \(\text{L}_*(\lambda, t) \approx \text{L}_*^\text{UV}(t) / \Delta \lambda_\text{UV}\), with \(\text{L}_*^\text{UV}\) the fraction of the bolometric luminosity that is due to UV radiations (see Appendix \ref{section:xuv}). We finally obtain:

\begin{equation}
    \text{t}_\text{ph}^{\text{f} \, *}(r, t)^{-1} = \frac{\text{L}_*^\text{UV}(t) \, \sigma_{\rm w}}{\text{h}\text{c} \, 4 \pi r^2}
\end{equation}

Then, we must calculate the column density of water along the line of sight from the inner part of the disc (at the distance \(r_\text{in}\) from the central star) to the distance \( r \):

\[
N_{\text{H}_2\text{O}} = \int_{r_{in}}^r \frac{\Sigma(R)}{\mu_{\text{H}_2\text{O}} \, \text{m}_\text{p} \, \text{H}_\text{z}(R)} \, dR
\]

\subsubsection{The interstellar radiation field contribution}

The interstellar radiation field (ISRF) spectrum \citep[based on ][]{draine_photoelectric_1978} is assumed to be constant in time.
The photons hit the disc from top and bottom. The optical depth is then much smaller than that of photons arriving from the central star. However, the ISRF luminosity is also low compared to the star's luminosity in the inner few au, but both contributions may be non-negligible.
For such a geometry, $N_{\text{H}_2\text{O}} \ \mu_{\text{H}_2\text{O}} \,  \text{m}_\text{p}$ is simply equal to the $\text{H}_2\text{O}$ surface density with $N_{\text{H}_2\text{O}} = {\Sigma_{\text{H}_2\text{O}}}/({\mu_{\text{H}_2\text{O}} \, \text{m}_\text{p}})$ .
One can calculate the free photodissociation timescale under the ISRF radiation only as follows \citep{rollins_chemical_2012}
\[\text{t}_\text{ph}^{\text{f}, \text{ISRF}} = \left(\int \text{I}(\lambda)_\text{ISRF} \sigma_\text{w} d \lambda\right)^{-1} \simeq 39  \ \text{yr}, \] where \(\text{I}(\lambda)_\text{ISRF}\) is the interstellar radiation field \citep[][]{draine_photoelectric_1978}.

\subsection{Ice sublimation} \label{sub_sec:ice_sublimation}

Our model merges and updates the two ice sublimation models described in \cite{kral_impact-free_2024} for water ice, and \cite{huet_late_2025} for CO. As in \cite{kral_molecular_2021}, we assume the gas density sublimation rate as a function of temperature $T$ to be given by
\begin{equation}
    \dot{\rho}(T) = S P_{\text{H}_2\text{O}} \sqrt{\frac{\mu_{\text{H}_2\text{O}}  \, \text{m}_\text{p}}{2\pi \, \text{k}_\text{B} \, T}},
    \label{eq:rho_dot}
\end{equation}
\noindent where $S = 3 (1-\Psi) / r_p$ is the total interstitial surface area of the pores of radius $r_p$ and porosity $\Psi$, $P_{\text{H}_2\text{O}}$ is the equilibrium pressure assumed to be given by the Clausius-Clapeyron law $P_{\text{H}_2\text{O}} = P_0 \ \exp{\left[\frac{\mu_{\text{H}_2\text{O}} \,  \text{m}_\text{p} \ h_{\rm sub}}{\text{k}_\text{B}}\left(\frac{1}{T_0} - \frac{1}{T}\right)\right]},$ where $P_0$ is the pressure at the reference point $T_0$, and $h_{\rm sub}$ the water sublimation enthalpy. 
We set $P_0 = 10^5$ N.m$^{-2}$, $T_0 = 373$ K, $ h_{\rm sub}=2.78 \times 10^6$ J.kg$^{-1}$ as in \cite{kral_impact-free_2024}. The asteroids are modelled as grey bodies (black bodies with an albedo A). We also use \(a_0\) as the distance at which the temperature is equal to \(T_0\) for a solar luminosity and a null albedo.
Then, the surface temperature of asteroids is given by:

\begin{equation}
    T_{\rm ast}(a, t) \sim T_0 \left(\frac{L_*(t)}{L_\odot}\right)^{1 /4} \ \left(\frac{a}{a_0}\right)^{-1/2} \ (1 - A) ^ {1 / 4},
    \label{eq:temp_ast}
\end{equation}
\noindent where $a$ is the asteroid's semi-major axis, $L_*$ is the central star's luminosity, which depends on time (see section \ref{sub_seq:luminosity}).

To get the temperature at a depth $d$ of an asteroid of size $s$, we have integrated the thermal diffusion equation in spherical coordinates

\begin{equation}
    \frac{\partial T}{\partial t} = K \left(\frac{\partial^2 T}{\partial d} - \frac{1}{(s-d)}\frac{\partial T}{\partial d}\right),
    \label{eq:thermal_diff}
\end{equation}

\noindent where $K$ is the thermal diffusion coefficient \citep{fourier_theorie_1822}. For asteroids, $K \sim 10^{-5} \ \mathrm{m^2 . s^{-1}}$ \citep{opeil_thermal_2010}.

At the surface, the temperature is set by Eq. \eqref{eq:temp_ast}. At the centre of the asteroid ($d=s/2$), the energy flux is assumed to be zero. We then integrate this equation for different asteroid sizes and distances from the central star\footnote{Since the temperature radial profile is known, there is no need to integrate this equation for all semi-major axes. It is integrated for an arbitrary radius $a_0$, and then the result at the radius $r$ is obtained by multiplying by $\left(\frac{r}{a_0}\right)^{-1/2}$.}. Using those temperatures, we can estimate the gas production rate for each asteroid of size $s$ at the distance $a$ from the central star via Eq. \eqref{eq:rho_dot}. Since the amount of ice available for sublimation, set by the initial ice to refractory mass ratio $f_{\rm ice} = 0.2$, similar to \cite{kral_impact-free_2024}, is limited, the gas production rate is set to zero when all ice sublimated.

We calculate a table of the surface density production rate $\dot{\Sigma}_{\rm belt}$ (see Section \ref{sub_seq:viscous_diff}) as a function of time and the distance to the central star, assuming the asteroid belt has an initial ice mass $M_{\rm ice}$, and a number density of objects $n(a,s)$, where $a$ is the distance to the central star and $s$ is the size of the asteroid (see Section \ref{sub_seq:ast_belt}).

The interpolated $\dot{\Sigma}_{\rm belt}$ is then used in the Eq. \eqref{eq:diff} in the calculation of $\dot{\Sigma}$ which is the sum of the contribution of gas production at the belt level ($\dot{\Sigma}_{\rm belt}$) and removal because of accretion ($\dot{\Sigma}_{\rm accr}$, see Section \ref{sub_sec:planetary_accr}).
For Eq. \eqref{eq:omega_i}, we have different expressions for $\omega_{\text{H}_2\text{O}}$ and $\omega_{H/O}$
\begin{equation}
    \begin{split}
        \dot{\omega}_{\text{H}_2\text{O}}^{\rm belt} &= \frac{\dot{\Sigma}_{\rm belt} ( 1 - \omega_{\text{H}_2\text{O}})}{\Sigma}, \\
        \dot{\omega}_{\rm H/O}^{\rm belt} &= -\frac{\dot{\Sigma}_{\rm belt} \omega_{\rm H/O}}{\Sigma},
    \end{split}
\end{equation}

\noindent where $\omega_{H/O}$ are the hydrogen and oxygen mass fractions, respectively. The terms \(\dot{\omega}_i^\text{belt}\) are the source (for water) and sink (for H and O) terms of Eq. \eqref{eq:omega_i} (see section \ref{sub_seq:viscous_diff}).

\subsection{A grid of stellar evolutionary tracks} \label{sub_seq:luminosity}

The evolution models used for this work were computed with the \cesamxx\footnote{\url{https://www.ias.u-psud.fr/cesam2k20/}} hydrostatic stellar evolution code (see \citealt{morel_cesam_1997,morel_cesam_2008,marques_seismic_2013,manchon_cesam2k20_2025}, for details on the numerical methods). We generated a small grid of models with an initial chemical composition that follows \citet{asplund_chemical_2021}'s solar determination, with opacity tables from the OPAL team \citep{rogers_rosseland_1992,iglesias_updated_1996}, adapted to these abundances. The equation of state (\eos) uses tables from the OPAL5 \eos  \,\citep{rogers_updated_2002}. The nuclear reaction rates follow the compilations from NACRE \citep{aikawa_nacre_2006} – except LUNA, \citealt{broggini_luna_2018}, for the ${}^{14}{\rm N(p},\gamma){}^{15}{\rm O}$ reaction. The convection is modelled with the mixing-length theory formulation of \citet{henyey_studies_1965} and the atmosphere is retrieved using the Hopf function $q(\tau)$ given in \citet{hubeny_theory_2014}. Effects of the atomic diffusion is not accounted for, while the rotation is described with the formalism of \citet{talon_rotational_1997}, where the evolution of the angular momentum distribution inside the stellar radiative zone is the result of the competing effect of the advection of angular momentum by meridional circulation and the diffusion of angular velocity by shear-induced turbulence. We set the initial distribution of angular momentum with the disc-locking model \citep{bouvier_angular_1997}, which assumes that the convective envelope of a star is magnetically coupled with its surrounding disc and forced to co-rotate with it with a period $P_{\rm disc}$ during the lifetime of the disc $\tau_{\rm disc}$. Angular momentum is lost by the star during its evolution through magnetized winds at the surface. This effect is modelled with the angular momentum loss prescription of \citet{matt_mass-dependence_2015}.

In order to build our grid, we started by computing a calibrated solar model that reproduces the actual luminosity (solar unit), $T_{\rm eff} =5772~\Tunit$ and surface angular velocity $\Omega_{\rm s} = 2.86\times10^{-6}~\omegaunit$ of the Sun, by tuning the initial He content $Y_0$, the mixing-length parameter $\alpha_{\rm MLT}$ and a quantity $T_{\rm wind}$ that controls the strength of the angular momentum loss at the surface \citep[see][]{matt_mass-dependence_2015}. We then computed 5 models with mass ranging from $0.7$ to $2.0$ M$_\odot$ with a step of $0.325 \ M_\odot$, a disc period $P_{\rm disc} = 10~\mbox{days}$ and a disc lifetime $\tau_{\rm disc} = 5~\mbox{Myrs}$. The calibrated value of $T_{\rm wind} = 1.62\times10^{31}~\mbox{erg}$ is kept the same for all our models. 

Our current understanding of the transport of angular momentum in stellar interiors is notoriously deficient \citep[e.g.][]{eggenberger_angular_2012,marques_seismic_2013,goupil_seismic_2013,cantiello_angular_2014,fuller_angular_2014,ouazzani__2019}, transport mechanisms are lacking and the interior profile of $\Omega$ shall not be trusted. However, only the surface angular velocity is relevant for this work and our calibration of $\Omega_{\rm s,\odot}$ ensures the correct trends and order of magnitude for its evolution through the star life. Moreover, the angular velocities considered here are small, and the rotation should have little effect on the stellar structure. From these rotating models, we could estimate the XUV flux with the scaling relations described in App. \ref{section:xuv}. .

\subsection{The simulation set} \label{sub_sec:simulation_set}

The objective of this paper is to conduct an exhaustive review of systems capable of hosting secondary water gas discs. For each configuration, we want to obtain the amount of water gas they can contain to guide future observations, as well as estimate the amount of water that can be accreted onto a hypothetical inner planet. To do this, we have chosen a wide parameter space to cover most plausible cases, and be able to draw clear general conclusions about exoplanetary systems. Hence, we have run 120 simulations to map every combination of our studied parameters as described in Table \ref{table:simulations}.

\begin{table*}
\caption{Parameters used for our set of 120 simulations.}
\label{table:simulations}      
\centering          
\begin{tabular}{ccccc}
\hline\hline       
                     
\(\text{M}_* \)\tablefootmark{a} \([\text{M}_\odot]\) & \(\text{M}_\text{belt}\)\tablefootmark{b} \([\text{M}_\oplus]\) & \(\text{t}_\text{init}\)\tablefootmark{c} [Myr] & \(\alpha\)\tablefootmark{d} & \(\text{a}_\text{belt} / \text{a}_\text{iceline}\)\tablefootmark{e}
\\ 
\hline                    
   0.7, 1, 1.35, 1.68, 2  & \(10^{-3} \), \(10^{-1}\) & 1, 5 & \(10^{-3} \), \(10^{-1}\) & 0.7, 1, 1.3 \\ 
\hline                  
\end{tabular}
\newline
\tablefoottext{a}{\(\text{M}_*\) is the central star's mass}
\tablefoottext{b}{\(\text{M}_\text{belt}\) is the total mass of the asteroid belt}
\tablefoottext{c}{\(\text{t}_\text{init}\) is the protoplanetary disc's lifetime, which sets the initial time of the simulation}
\tablefoottext{d}{\(\alpha\) is the viscous parameter}
\tablefoottext{e}{\(a_\text{belt}\) the belt's mean semi-major axis and \(a_\text{iceline}\) the iceline location at the beginning of the simulation.}
\end{table*}

The first parameter we vary is the mass of the central star $\text{M}_*$. We selected five masses roughly linearly spaced in the range 0.7 to 2 solar masses (0.7, 1, 1.35, 1.68, and 2 solar masses corresponding to spectral types K, G, F, A7, A2, respectively). 

Another parameter is the dissipation time of the protoplanetary disc, which is important as it sets the initial conditions for our simulations. However, it is not well constrained for such a wide range of star masses, but happens within a few million years \citep{ribas_disk_2014, ribas_protoplanetary_2015, polnitzky_precise_2026}.
Different values could, however, lead to some major differences in our simulation results because the stellar luminosity varies rapidly between zero and ten million years (see Figure \ref{fig:bol_lum}). To account for different plausible cases, we took two initial times \(\text{t}_\text{init}\) of one (early) and five million years (late), respectively. 

It is interesting to note that for all stars studied here, they exhibit a luminosity peak following a rapid decline early in the star's life. After the peak, the luminosity then decreases slightly to reach a steady-state value over the duration of our simulations. The magnitude of this peak increases with the mass of the star, happening progressively earlier in time.

For instance, for stars with a mass of 0.7~M$_\odot$, the luminosity at the beginning of the simulation (for both \(\text{t}_\text{init}=1\)~Myr and 5~Myr) is significantly higher than the maximum luminosity at the time of the rebound, nearly 50~million years later. This is also the case for solar-type stars, at least for simulation with \(\text{t}_\text{init}=1\)~Myr. In all other cases, the initial luminosity is lower than that at the peak. For stars with masses of 2~M$_\odot$, the rebound occurs at 5~Myr, which also corresponds to one of the tested \(\text{t}_\text{init}\) values.
In general, we expect simulations with an initial time of five million years to be more sensitive to luminosity peaks, such as the one at 20 Myr for a solar-type star, \citep{kral_impact-free_2024} since the initial luminosity is smaller than in the early configuration.

\begin{figure}[h]
\begin{centering}
\includegraphics[scale=0.5]{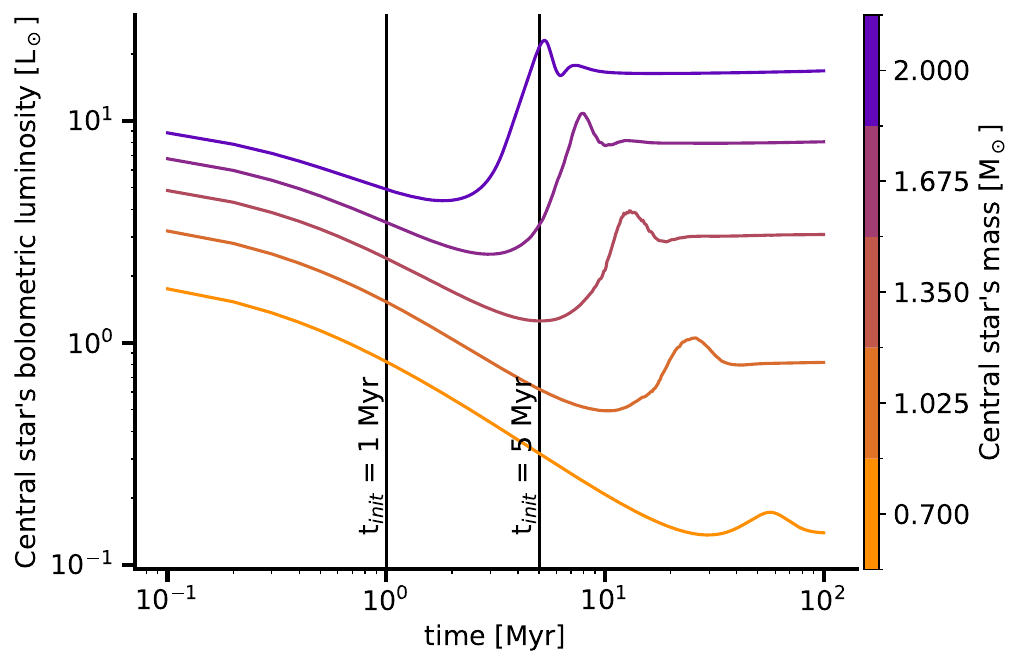}
\caption{\emph{Stellar bolometric luminosity for the different star masses tested in this study (in colour) as a function of time (see Section \ref{sub_seq:luminosity}). The two vertical black lines delimitate the two initial times \(\text{t}_\text{init}\) tested, 1 and 5 Myr.}}
\label{fig:bol_lum}
\end{centering}
\end{figure}

As in \cite{kral_impact-free_2024}, we model two initial belt masses. Our lightest belt is similar to the modern asteroid belt in the Solar System, with a mass of $M_{\rm belt} = 10^{-3} M_\oplus$. The other belt has a mass of $0.1 \ M_\oplus$, and could be dispersed later, as may have happened in our Solar System \citep[see e.g., ][]{clement_mars_2018, clement_excitation_2019}.

Moreover, we assume that the exo-asteroid belts are close to the water ice-line when the protoplanetary disc dissipates. We have modelled three different locations for the mean semi-major axis of the belt {\(\text{a}_\text{belt}\), one at the ice-line itself \(\text{a}_\text{iceline}\), one at $0.7 \ \text{a}_\text{iceline}$, and one at $1.3 \ \text{a}_\text{iceline}$ (see Section \ref{sub_seq:ast_belt} for the details of the calculations involved to find the ice-line location). 

Lastly, we use two values for the viscous parameter $\alpha$. It is not well constrained from observations in debris discs \citep{kral_imaging_2019}. However, \cite{kral_magnetorotational_2016,cui_dynamics_2024} provide some estimates for the viscous parameter using analytical models of angular momentum transport in discs. They find that given the high ionisation and low gas density, $\alpha$ could be relatively high compared to protoplanetary discs due to magnetohydrodynamic instabilities and molecular viscosity.

Therefore, we test two plausible values $\alpha=10^{-3}$ and $\alpha=10^{-1}$.
In all simulations, we have added a putative inner planet with a semi-major axis $a_p=0.8 \ \text{a}_\text{belt}$. We take a planet with an arbitrarily high mass such that the ratio \(\text{R}_\text{H} / \text{H}_\text{z} > 1 \) (see Section \ref{sub_sec:planetary_accr}). We will later discuss the influence of the planet's mass on the accreted water mass.

\section{Results}

\subsection{A young solar-system-like asteroid belt as a reference} \label{sub_seq:fiducial_result}
We choose to take a solar-system-like simulation as a reference to compare our results with \cite{kral_impact-free_2024}, and to then describe the key elements common to all simulations. In the reference simulation we take the following parameters: M$_* = 1$~M$_\odot$, M$_\text{belt} = 0.1$~M$_\oplus$, t$_\text{init} = 5$~Myr, $\alpha = 0.1$, and a$_\text{belt} =$~a$_\text{iceline}$. This simulation is therefore the closest representation of the young solar system in our simulation set, and is similar to the model of \cite{kral_impact-free_2024}.

   \begin{figure}[h]
\begin{centering}
\includegraphics[scale=0.5]{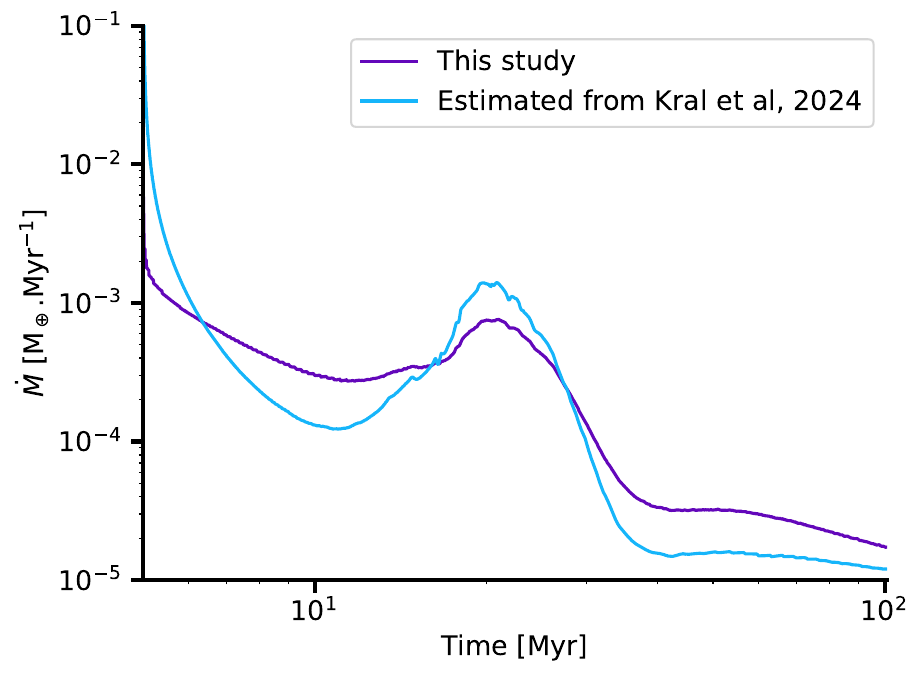}
\caption{\emph{Gas mass production rate as a function of time for the reference simulation (in purple) and the model described in \cite{kral_impact-free_2024} (in blue).}}
\label{fig:fid_mdot}
\end{centering}
\end{figure}

As in the latter study, we find that there are two distinct phases of gas production due to the central star's luminosity rebound at about 20~Myr (see Figure~\ref{fig:fid_mdot}). However, the amplitude of the secondary mass production rate is found to be lower than that in \cite{kral_impact-free_2024} as shown in Figure~\ref{fig:fid_mdot}. It is because of thermal diffusion, which is a new ingredient added in our simulations. Even though thermal diffusion has a negligible effect on the total mass of gas produced over the lifetime of the disc, it smooths out the mass production rate. Thus, instead of having two distinct production peaks with negligible gas production in between, we observe a more constant gas production. Although two production peaks are still distinguishable (one at the beginning of the simulation and another 20~million years later), their intensities are weaker than in the model of \cite{kral_impact-free_2024}. An interesting consequence is that by producing gas more slowly, the critical gas surface densities required for shielding against photodissociation might be harder to reach. We note that, similar to \cite{kral_impact-free_2024}, the gas mass produced is dominated by the gas released during the rebound phase.

For this reference simulation, we find that the accreted mass onto the inner Earth-mass planet is approximately $5 \times 10^{-3}$~M$_\oplus$, similar to the accreted mass on Earth in \cite{kral_impact-free_2024}. Therefore, the improved gas production model presented here has a minor impact on the total amount of gas produced, hence on the accreted water mass onto the inner planets, but some impact on the gas surface density, which is important for photodissociation and observations in general.

   \begin{figure}[h]
    \begin{centering}
    \includegraphics[scale=0.4]{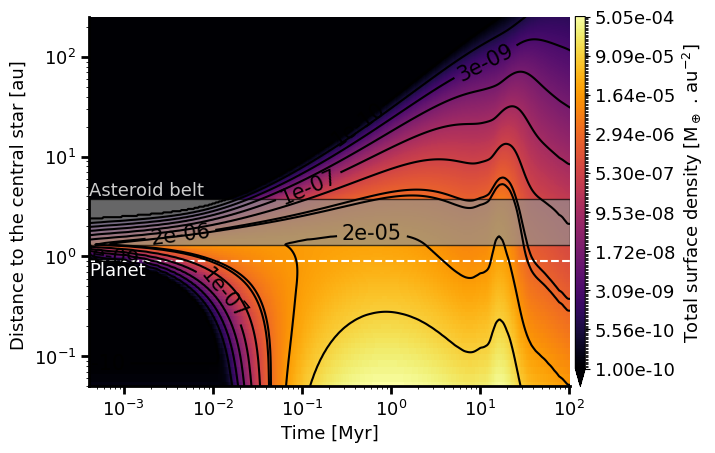}
    \includegraphics[scale=0.4]{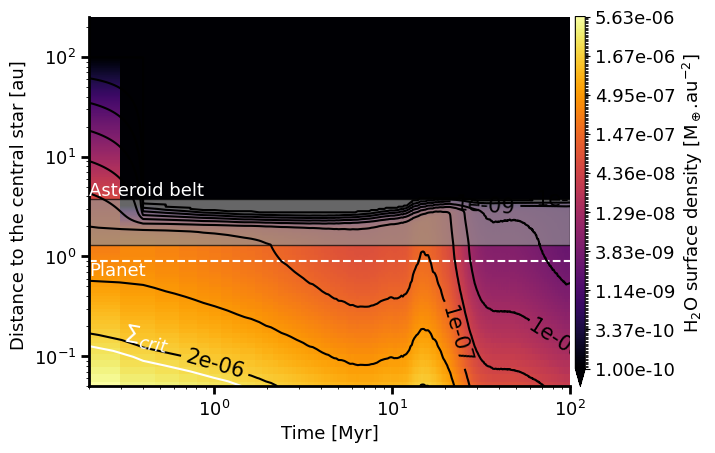}
    \caption{\emph{Reference simulation: Total surface density of the gas (top) and that of water vapour only (bottom) as a function of the distance to the central star and time. The white level line corresponds to the critical surface density $\Sigma_\text{crit}$. The grey shaded area represents the belt. The dashed line is at the planet's location}.}
    \label{fig:fid_sigmas}
    \end{centering}
    \end{figure}

In the reference simulation, we find that the water surface density is always smaller than the critical surface density required for shielding, both in the radial and vertical directions (except in the very inner disc at the very beginning of the simulation). This is because the gas production rate is never high enough to produce a massive secondary water gas disc in this simulation; instead, all the water is quickly photodissociated before it can reach the critical surface density (see Figure~\ref{fig:fid_sigmas}).

Although the critical surface density for water is never reached, there is still a small portion of water vapour in the disc, which is due to water that is constantly released from the belt before it gets photodissociated. We find that the water mass fraction is low, less than 10 \%, compared to the mass fraction of neutral oxygen and hydrogen. The surface densities reach on the order of $\Sigma_\text{crit} / 10$, even at the time of the luminosity rebound, 20~million years after the beginning of the simulation, where it could be expected to peak. We observe the same behaviour in the simulation with a viscosity of \(\alpha= 10^{-3}\), which is 10 times lower than that used in \cite{kral_impact-free_2024}. The decrease in viscosity does not lead to a sufficient increase in surface density to provide long-lasting protection against photodissociation.

\subsection{Study of secondary gas discs produced from exo-asteroid belts}
\subsubsection{Efficiency of gas production} \label{sub_seq:efficiency_production}
The gas production efficiency which is the total amount of gas released, compared to the maximum amount of gas that can be produced by the belt (equal to \(f_{\text{ice}} M_{\text{belt}}\)), varies significantly across all simulations.

   \begin{figure}[h]
\begin{centering}
\includegraphics[scale=0.5]{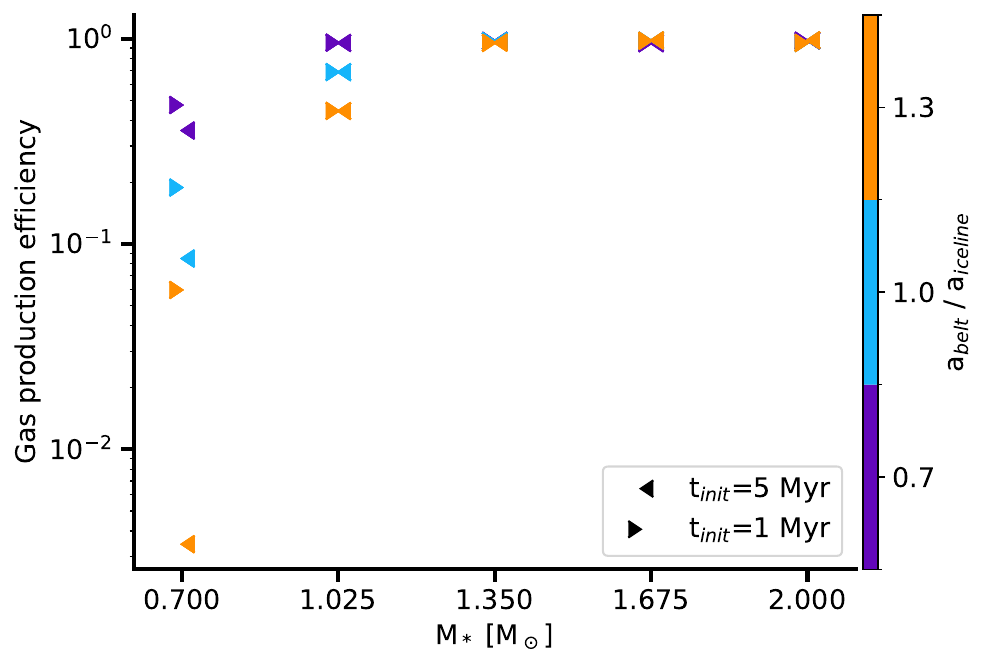}
\caption{\emph{Gas production efficiency (\(\text{M}_\text{gas produced}\) / \((f_\text{ice} \ \text{M}_{belt})\)) as a function of the central star's mass for all simulations. The purple triangles are for the belts located at a factor 0.7 of the ice-line, the blue triangles are for the belts centred around the ice-line, and the orange dots are for the belts 1.3 times farther than the ice-line. The right and left triangles correspond to simulations that start 1~Myr and 5~Myr after the birth of the system, respectively.}}
\label{fig:diag_mprod}
\end{centering}
\end{figure}

For stars with masses beyond 1~M$_\odot$, we find that the mass production efficiency is always higher than 95\%, even for simulations with belts farthest from the star. For lower-mass stars, the mass production efficiency can drop as low as \(3 \times 10^{-2}\) (see Figure~\ref{fig:diag_mprod}). In absolute terms, the luminosity increases significantly with the mass of the stars (see Figure~\ref{fig:bol_lum}). However, this does not necessarily result in more efficient sublimation, as the locations of the belts are chosen based on the ice-line at the time of the protoplanetary disc dissipation. Rather, this highly efficient gas production around more massive stars is explained by a luminosity rebound amplitude – which can happen very early for A-type stars (see Fig.~\ref{fig:bol_lum}) – that is much greater than the initial luminosity, regardless of \(t_{\text{init}}\). In these cases, even the most distant bodies are heated sufficiently to release a significant fraction of their ice.

For stars with masses equal to or less than one solar mass, this is no longer the case. The luminosity of the rebound is lower than the initial luminosity. The location of the belt relative to the ice-line then becomes the main factor influencing gas production through sublimation of the ice present in the asteroids (see Figure~\ref{fig:diag_mprod}). For the closest, warmest belts, the gas production efficiency never drops below 0.3, two orders of magnitude higher than the gas production efficiency for the farthest belts. For these stars, the last parameter that influences the gas production efficiency is the initial time of the simulation (which matches the dissipation timescale of the protoplanetary disc). We find a lower gas production efficiency for an initial time of 5~Myr. This is expected given the evolution of the luminosity curves: the luminosity is much higher from 1 to 5~Myr than afterward for these stars (see Figure~\ref{fig:bol_lum}).

\subsubsection{Maximum masses of the gas discs} \label{sub_seq:maximum_mass}
With the same total mass in an exo-asteroid belt, very different kinds of gas discs can be produced. For instance, considering an initial belt mass of \(10^{-3}\)~M$_\oplus$, we find maximum gas disc masses ranging from \(\sim 10^{-8}\)~M$_\oplus$ for high-viscosity discs around low-mass stars to \(\sim 10^{-4}\)~M$_\oplus$ for discs around 2~M$_\odot$ stars. To be able to detect these gas discs, it is critical to determine how massive they can be and at what time the maximum mass is reached, as a function of the parameters considered in this study.

Viscosity can significantly decrease the maximum disc mass by almost one order of magnitude with a viscous parameter of 0.1 compared to \(10^{-3}\) (see Figure \ref{fig:diag_mmax}). The belts closest to their stars produce the most massive discs. As the belt is getting closer to the central star, the gas production rate increases faster than the gas dispersion rate related to the viscous timescale. From the outermost to the innermost belt, the viscous diffusion timescale decreases by approximately 60\%. Meanwhile, the characteristic sublimation timescale increases by at least a factor of 300, as the belt is closer to the star, which increases the mass of the disc at the time of the production peak. Given a fixed star mass, we observe up to one order of magnitude difference between the maximum mass of a disc produced by a belt located 30\% closer than the ice-line compared to a disc located 30\% farther away. No difference is observed in the maximum mass of the discs as a function of the belt distance for more massive stars. This can be explained by the fact that the intensity of the luminosity rebound (see Figure~\ref{fig:bol_lum}) is greater than the initial intensity. The iceline is thus pushed further out into the system during the rebound, enabling the sublimation of ices that were initially too cold, regardless of the belt position chosen in our simulations.

Finally, for stars with masses equal to or less than one solar mass, the maximum mass of the disc is significantly influenced by the initial time \(t_{\text{init}}\) at which the debris disc simulation starts. The maximum mass of the disc can be up to one order of magnitude lower for simulations starting at 5~Myr compared to those starting at 1~Myr. This is a direct consequence of lower gas production for simulations starting later around low-mass stars (see Figure~\ref{fig:diag_mprod}), due to the evolution of luminosity, which decreases by almost two between 1 and 5~Myr and never rises as high during the luminosity rebound (see Figure~\ref{fig:bol_lum}).

It is also important to determine when this maximum mass is reached. The gas production can be divided into two phases: a first burst of gas production while the young star's luminosity decreases progressively, and a second burst (if there is ice left) at the luminosity rebound (see Figure~\ref{fig:bol_lum}). The solar-type stars experience the strongest second burst of gas, since after 10~Myr, the gas mass left to be produced is at least 35\% of the total gas mass produced, and can reach 85\% (see Figure~\ref{fig:diag_mprod_time}). For stars with masses between 1.675 and 2~M$_\odot$, there is no secondary burst since there is less than 10\% of gas left to be produced after 10~Myr (see Figure~\ref{fig:diag_mprod_time}). This can be explained by the very early luminosity rebound of those stars. Most of the gas is produced during the rebound, when the ice line moves outward through the system, sublimating ice that was initially too cold. However, since the rebound occurs very early on, most of the gas is produced at the beginning of the simulation. For the low-mass stars (0.7~M$_\odot$), we find that the mass fraction left to be produced after 10~Myr is up to one order of magnitude higher and at least twice as much for simulations that start 5~Myr after the system formation instead of 1~Myr (see Figure~\ref{fig:diag_mprod_time}).

\subsubsection{The maximum water surface density} \label{sub_seq:maximum_water_sd}
The total disc mass is composed of both molecular water and its photodissociation products, H and O. Since water gas can be observed in the infrared and sub-millimeter wavelengths, it is important to consider how it is affected by photodissociation. We find that photodissociation has a major effect in all simulations, and the discs are never globally shielded over long timescales from interstellar and stellar radiation (except for very massive gas discs produced either by a heavier belt or a lower viscosity, see section \ref{sub_seq:water_obs}). We note that this is the frame of our 1D model, which does not account for 2D or 3D effects, which may complicate the picture. We find that the maximum water disc mass is reached very early on in the simulations, always before 1.3~Myr.

This is because the gas generation rate is very high early on in the simulations due to the higher stellar luminosity, together with small grains releasing a significant fraction of their ice quickly as thermal diffusion acts very fast. The critical surface density required for self-shielding against stellar and interstellar radiation is then reached for some million years, allowing the existence of secondary water discs with masses up to \(7 \times 10^{-3}\)~M$_\oplus$ (see Figure~\ref{fig:diag_mwater_twater}).

   \begin{figure}[h]
\begin{centering}
\includegraphics[scale=0.5]{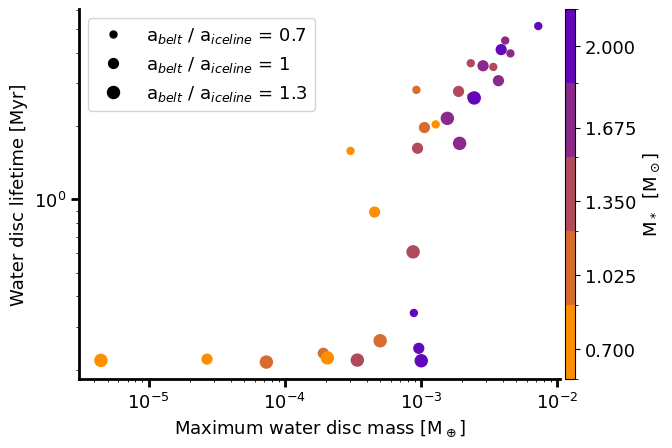}
\caption{\emph{Lifetime of water discs, defined as the time required for the disc mass to decrease by a factor of ten relative to the maximum mass, as a function of the maximum mass of the water disc. The colours correspond to the mass of the central star, and the size of the points indicates the distance of the belt relative to the iceline (from smallest for the closest belts to largest for the most distant belts).
}}
\label{fig:diag_water_disc_lifetime}
\end{centering}
\end{figure}

We observe a population of discs reaching their maximum mass around 1~million years after the start of the simulation, while all other simulations reach their maximum mass earlier, after around 200,000 years. This longer lifetime is explained by a longer gas viscous timescale at low $\alpha$ values, which allows higher surface densities to be maintained, reaching values above the critical density that protects against photodissociation. Only the biggest belts with a total mass of 0.1 \(\text{M}_\oplus\) can be shielded from photodissociation.  
For systems with the most massive belts (0.1 Earth mass) and the lowest viscosity, we observe that the critical mass for a water disc to be shielded from photodissociation is on the order of \(10^{-3}\) Earth mass. The typical lifetime of water discs, defined as the time required for the maximum disc mass to decrease by a factor of 10, is on the order of a few million years (see Figure~\ref{fig:diag_water_disc_lifetime}). We find that, for systems around massive stars, all the discs considered are shielded from photodissociation, whereas for systems around less massive stars, only the discs produced by belts closest to the central star are shielded from photodissociation and survive for several million years. The maximum lifetime of water gas discs is then close to the viscous diffusion timescale (see Section~\ref{sub_seq:viscous_timescale}). The gas disc, initially shielded from photodissociation, spreads until the surface density falls below the critical density, at which point the gas gets photodissociated.

\subsection{Accretion by the inner planet} \label{sub_seq:res_accr}

   The point of this study is not to provide a precise estimate of the accreted water mass onto planets for a specific system but rather provide estimates as a function of the central star, the belt and the planet's masses.
   Hence we have fixed a specific architecture and we will now discuss on the potential modulations for all of those parameters.
In each of our simulations, we placed an inner planet at the position \(a_{\rm p} = 0.8 \, a_\text{belt}\) (see Section~\ref{sub_sec:simulation_set}). We do not consider the addition of more planets which could influence the incoming flux onto the planet (but see later in this subsection). While the planet's position may affect the local surface density of the disc, it only influences the accreted mass through the factor \(\text{R}_\text{H} / \text{H}_\text{z}\) (see Section~\ref{sub_sec:planetary_accr}, the Appendix~\ref{sub_seq:rh_H} and Equation~\ref{eq:rh_H}). Placing the inner planet 0.8 times closer than \(a_\text{belt}\) results in a negligible increase in the \(\text{R}_\text{H} / \text{H}_\text{z}\) factor of about 1.06 compared to an hypothetical planet located at  \(a_\text{belt}\). A planet placed ten times closer would see its \(\text{R}_\text{H} / \text{H}_\text{z}\) factor increase by approximately 1.78.

In the case of an Earth-mass planet orbiting in a system with its belt located at the iceline, the \(\text{R}_\text{H} / \text{H}_\text{z}\) factor is always less than 1, but greater than 0.7, regardless of the mass of the star considered\footnote{Similar to all our simulations, we take a temperature radial power law \(\alpha_T = -0.5\) and a composition of pure water for the gas.} (see Figure \ref{fig:rh_h}). Taking into account the location of the belt, any planet with a mass greater than 3.62 Earth masses will have a \(\text{R}_\text{H} / \text{H}_\text{z}\) ratio greater than 1 and will thus accrete the maximum possible mass. This mass threshold drops to 2.91 Earth masses if the belt is at the iceline and to 2.82 Earth masses if it is 30\% closer. A planet as low in mass as Mars (with a mass on the order of 0.1 Earth mass) will accrete approximately 0.46 times the mass that an Earth-mass planet could accrete under equivalent conditions.

As a reminder, we arbitrarily set the \(\text{R}_\text{H} / \text{H}_\text{z}\) factor equal to 1 in our simulations. We observe that the mass accreted onto the planets is close to \(\text{f}_\text{accr}\) times the total mass of gas produced, with fluctuations that we can attribute to our integration method rather than a physical phenomenon (see Appendix~\ref{sub_seq:f_evolution}). Since Eq.~\eqref{eq:diff} does not favour any direction of propagation over the other, one would expect, in principle, that as much mass flows outward as inward in the system. This implies a maximum accreted mass twice as low as that observed for systems with outers planets that may accrete all of the gas mass that flows outward. While we can confirm that initially as much mass flows inward and outward from the system, we also observe that when mass production stops, the direction of the flux reverses, and the gas initially diffused outward subsequently diffuses inward (see Section~\ref{sub_seq:f_evolution}). This explains the high accreted masses observed and also accounts for the slight dispersion in our calculations of accreted masses.

Indeed, our boundary condition on the flux at the outer edge imposes that no mass flux can enter from the outside. Any mass leaving our integration domain is definitively lost and cannot rediffuse inward as it would if the integration domain were larger. The simulations in which accretion is less efficient are those with a slightly shorter spatial integration domain. We defined the outer limit of our integration domain as one hundred times the mean radius of the belt, \(a_\text{belt}\). However (see Section~\ref{sub_seq:viscous_timescale}), the characteristic viscous diffusion timescale depends linearly on distance. We therefore expect that simulations with belts closest to their central stars, particularly belts around M stars, will have a more significant outward mass flux lost that can never be recovered.

We thus estimate that the mass accreted onto the planets is equal to \(\text{f}_\text{accr}\) modulated by variations in the \(\text{R}_\text{H} / \text{H}_\text{z}\) factor. Consequently, another outer planet is likely to reduce the amount of water accreted onto inner planets by intercepting the mass flux initially flowing outward. In the most unfavourable case, one can therefore expect the mass accreted onto inner planets to be half as much as without the outer planet.

Furthermore, this implies that planets around more massive stars will accrete more gas than planets around less massive stars, as the efficiency of gas production is lower in the latter (see Section~\ref{sub_seq:efficiency_production}). This effect will be particularly pronounced for planets in systems with belts further from the central star. The accreted mass would then be on the order of one per cent of the initial ice mass in the belt and thus on the order of one-tenth of a per cent of the total belt mass. Conversely, for systems around F- and A-type stars, the accreted mass will be on the order of one-tenth of the total belt mass or up to 0.01 Earth mass in the case of simulations explored, though more massive belts could exist (see Section~\ref{sub_seq:water_obs}). One can therefore  consider accreted masses on the order of one-tenth of an Earth mass in favourable cases (See Figure \ref{fig:water_accr_mass}).

\begin{figure}[h]
\begin{centering}
\includegraphics[scale=0.5]{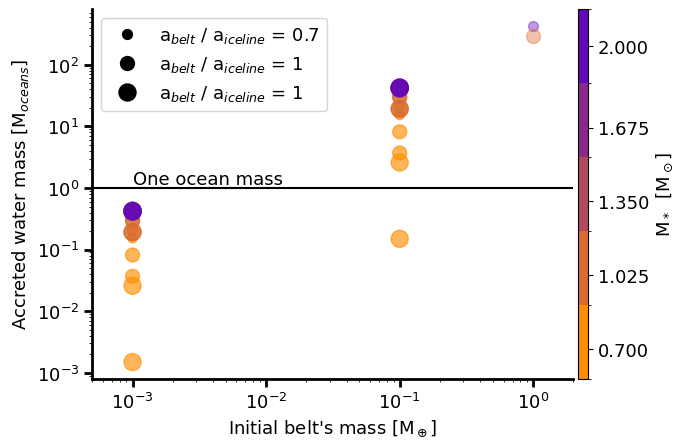}
\caption{\emph{Accreted water mass (in Earth ocean masses) onto the inner planet as a function of the initial belt's mass and the central star mass. The two dots at one Earth mass correspond to dedicated extra simulations (see Section \ref{sub_sec:evoltion_water_disc} for a description of extra high mass simulations with belts of 1 M$_\oplus$).}}
\label{fig:water_accr_mass}
\end{centering}
\end{figure}

\section{Discussion}

\subsection{The sublimation process}

The addition of a proper treatment of photodissociation to our model, coupled to the modelling of a larger number of exoplanetary systems, called for an improvement of the gas generation model compared to that of \cite{kral_impact-free_2024}. While their model used the sublimation rate as the characteristic timescale, and assumed that sublimation occurred on timescales much faster than thermal diffusion, our model also includes thermal diffusion within the solids.

The main effect is to smooth the gas production rate over time, as the high temperatures required for sublimation, induced by heating from the central star, are reached increasingly later at the core of bodies as they grow. However, the total gas masses produced remain largely equivalent. This new addition to the model is particularly relevant when including photodissociation, as a lower production rate, consequence of the thermal diffusion, would prevent the critical surface density for water photodissociation from being reached.

However, since sublimation occurs at depth, the gas takes a non-zero amount of time to diffuse through the refractory porous material of asteroids. The characteristic time associated with this process is \citep[e.g.][]{kral_molecular_2021}
\[
\tau_{\rm diff} = \frac{3}{4} (\Delta p)^2 \frac{\sqrt{2 \pi m_{\text{H}_2\text{O}}}}{\Psi r_p \sqrt{\text{k}_\text{B} T}},
\]
where \(\Delta p\) is the depth within the asteroid, \(r_p \sim 1 \, \mathrm{\mu m}\) is the typical pore radius \citep[see ][]{sultana_proprietes_2021}, \(\Psi\) is the porosity \citep[of the order of 0.6;][]{kral_molecular_2021}, and \(T\) is the temperature at depth \(\Delta p\). Let us compare this timescale to that of thermal diffusion. The characteristic time for thermal diffusion has the same \((\Delta p)^2\) dependence:
\[
\tau_{\rm therm} = \frac{(\Delta p)^2}{K},
\]
where \(K\) is the thermal diffusivity (in our model, \(K \sim 10^{-5} \, \mathrm{m^2 \, s^{-1}}\); see Section~\ref{sub_sec:ice_sublimation}). Thus, the ratio of the two timescales is
\[
\frac{\tau_{\rm diff}}{\tau_{\rm therm}} = \frac{3}{4} K \frac{\sqrt{2 \pi m_{\text{H}_2\text{O}}}}{\Psi r_p \sqrt{\text{k}_\text{B} T}} \sim 10^{-1} \quad \text{for} \quad T \sim 100 \, \text{K}.
\]
Therefore, it appears reasonable to neglect the molecular diffusion timescale compared to that of thermal diffusion in our model.

The second characteristic time in our gas production model is given by the sublimation rate (see Section~\ref{sub_sec:ice_sublimation}). To determine whether the molecular diffusion timescale is also negligible compared to that of sublimation, we calculate the ratio of the two timescales:
\[
\frac{\tau_{\rm sub}}{\tau_{\rm diff}} = \frac{4}{3 S P_{\text{H}_2\text{O}}} \frac{\rho_{ice} \Psi r_p \text{k}_\text{B} T}{\sqrt{2 \pi} m_{CO}} \frac{1}{(\Delta p)^2}.
\]
This time, the ratio depends on the depth \((\Delta p)^2\). The critical depth at which this ratio equals 1 is 78 km. This implies that for larger bodies, one should, in principle, also account for molecular diffusion in the model. However, at such depths, the characteristic thermal diffusion time exceeds 20 Myr. This implies that, in practice, a noticeable effect would only occur for temperature variations on timescales exceeding 20 Myr, approximately one-fifth of the total duration of our simulations. The typical variations in the luminosity of stars are much faster, sometimes operating over less than 1 Myr, which justifies the omission of molecular diffusion in our model

Like the gas model of \citet{kral_impact-free_2024}, our model includes free parameters.
The first is the initial mass fraction of ice, which is set to \(20\)~\%.
This fraction may vary from system to system, yet such variations do not fundamentally alter our conclusions.
Indeed, an increase (or decrease) in this ratio would simply be equivalent to an increase (or decrease) in the total mass of the belt in our model which is a parameter we already vary.
This uncertainty should therefore be considered when examining specific systems where the total mass of the belt is known.

In addition to this initial mass fraction of ice, we introduce a dependence on three key parameters.
The first is the thermal conductivity \(K\), representative in this study of the typical thermal conductivity of asteroids.
A higher thermal conductivity would imply faster thermal diffusion, reducing the significance of the smoothing effect observed in the production rate—most notably in Figure~\ref{fig:fid_mdot}.
The upper limit would thus be the value obtained by the model of \citet{kral_impact-free_2024}.
Conversely, a lower thermal conductivity would amplify this smoothing effect.
It should be noted that, while the smoothing effect is not negligible, it is secondary, as it does not affect the total mass of gas produced.
To achieve a shielded disc, the most critical factors are a massive belt and a significant increase in the luminosity of the central star, shifting the ice line well beyond the belt.
Since the variations in production rate between our model and that of \citet{kral_impact-free_2024} are less than an order of magnitude, we can reasonably assume that this uncertainty in \(K\) does not strongly influence our conclusions regarding the nature of the discs.

   The other two free parameters are the porosity and the pore size of the solids.
These parameters allow us to set the effective sublimation surface area within the solids.
This is an improvement over the model of \citet{kral_impact-free_2024}, which assumed sublimation only at the surface and not throughout the entire volume.
The surface area—and thus the production rate—is directly proportional to the porosity \(\Phi\).
Although porosity may also vary from body to body, we do not expect variations of several orders of magnitude, and thus no significant impact on our results.
We have deliberately chosen a low estimate for the pore size, leading to the highest possible sublimation rates.
If the minimum pore size were larger, the production rate would, in principle, decrease accordingly.
However, we must not overlook that the thermal diffusion timescale is often predominant over the sublimation timescale in the parameter space we explore, and it is this timescale that, in practice, modulates the gas production rate.
Thus, this uncertainty in pore size does not necessarily translate into uncertainty in the production rate, limiting the impact of this parameter on our results.

Our new model therefore shows no significant differences from the results produced by the model of \citet{kral_impact-free_2024}, while remaining more relevant in the context of a photodissociation study.
Indeed, although we introduce a few free parameters, we simultaneously lift structural assumptions made by \citet{kral_impact-free_2024}.

\subsection{Photodissociation from the central star}

While photodissociation by the central star is often neglected in cold debris disc models \citep[e.g.,][]{marino_population_2020, huet_late_2025}, it may be important to consider when modelling water because asteroid belts are much closer to the central star (typically 10 times closer than cold belts; see, for example, \citealt{matra_resolved_2025}), resulting in a stellar flux that is $\sim$100 times more intense.

The main argument for neglecting photodissociation by the central star is the optical depth of the gas disc, as it quickly becomes very high radially. The critical densities required to shield against photodissociation are therefore achieved rapidly. However, this argument only holds for discs that had time to viscously spread. Moreover, when considering relatively low gas production rates, this additional source of photodissociation may play a role in determining whether a water disc is shielded against stellar radiation.

In practice, for the simulations we carried out, we estimate that the effect of photodissociation by the central star is negligible compared to that caused by interstellar radiation. To check that, we conducted two complementary simulations based on our reference simulation, including only interstellar radiation or only stellar radiation. The simulation with only interstellar radiation produced a water disc similar to that observed in the reference simulation, with surface densities of water well below the critical density. Conversely, in the simulation with only stellar radiation, we observed a massive water disc, nearly as massive as the total gas disc (see Figure~\ref{fig:test_discs_masses}).

\subsection{Evolution of the water gas disc and planets}\label{sub_sec:evoltion_water_disc}

In order to prepare for future observations, it is important to study the mass evolution of water discs. Looking at our simulations, we find that, regardless of the systems considered, the mass of the water disc will be at its maximum no more than $\sim 1.5$ million years after the dissipation of the protoplanetary disc. The typical water disc lifetime (i.e. the time required for the disc mass to decrease by a factor of ten relative to the maximum mass) exceeds several million years, increasing the probability of detection (see Section \ref{sub_seq:maximum_mass}, \ref{sub_seq:water_obs}). In addition, even long after the maximum mass peak, the rate of water production may remain sufficient for the disc to be detectable for tens of millions of years (see Section \ref{sub_seq:water_obs} and Figure \ref{fig:flux_jwst_time}).
The viscous parameter is not well constrained in secondary water gas discs. Here, we have considered relatively high values of $\alpha$ based on typical values used for secondary CO gas discs \citep{kral_magnetorotational_2016, cui_dynamics_2024}, but water gas discs may be less ionised, thus reducing the effectiveness of instabilities that may be able to transport angular momentum \citep{kral_magnetorotational_2016,cui_dynamics_2024}. 

For a viscosity as low as \(10^{-5}\), the viscous timescale is 100 times higher than in previous low-viscosity simulations we explored (see Section \ref{sub_seq:viscous_timescale}). Such low $\alpha$ values could significantly increase the probability of detecting secondary water gas discs. A lower viscosity might also increase the local surface density, allowing the shielding from interstellar radiation of lower mass discs. To check that, we have then run a special simulation similar to the reference simulation but with a lower viscosity of \(\alpha = 10^{-5}\).
The maximum water surface density in this simulation increases by a factor 10 compared to the reference simulation, allowing a shielding from interstellar radiations contrary to the reference simulation. This maximum mass is reached 1.5 million years after the beginning of the simulation, and the typical lifetime for this water gas disc is about 7.45 Myr. To be complete, we also ran a simulation similar to the reference simulation but with a very low viscosity of \(\alpha = 10^{-5}\), and a belt mass ten times higher (\(\text{M}_\text{belt} = 1 \ \text{M}_\oplus\)). For this simulation, the maximum water disc mass is 20 times higher than the previous one and 200 times higher than the reference simulation. This maximum is reached later than for the simulation with \(\text{M}_\text{belt} = 0.1 \ \text{M}_\oplus\), 3.2 millions year after the beginning of the simulation. The water disc typical lifetime exceeds 10 millions years in this case. This very high surface density of gas over long timescales might also have an impact onto planet's migration. Using the estimated planetary migration rate from \cite{kral_impact-free_2024}, we can conclude that it is negligible for an Earth-like planet (or even a more massive planet) in the case of the reference simulation. For the more extreme case of an 0.1 Earth mass belt, we can estimate the migration of a planet with a mass five times that of Earth to be of the order of an astronomical unit. As the location of the planet is critical in triggering a runaway greenhouse effect \citep{turbet_runaway_2019}, migration induced by the secondary gas disc could significantly alter the planet's climate.

We find that when ice is present in the belt, and the stellar mass exceeds one solar mass in our simulations (F and A stars), it is likely that the belt releases gas at a significant rate. Indeed, 90\% of the ices present in belts around such stars sublimate within 10 million years. In this case, the maximum mass of the water disc is then proportional to the mass of the asteroid belt, reaching up to 50\% of the initially present ice mass for discs around 2-solar-mass stars, provided the gas viscosity is low.

For M- or G-type stars, the distance of the belt from the central star is crucial to determine the maximum water disc mass. Closer belts enable significant gas production. For solar-type stars, gas production is expected to be half as much for the most distant belts. For M-type stars, this production can be up to 10 times lower.  However, the light belts produced in those cases might still be detectable (see Section \ref{sub_seq:water_obs}) but for systems closer to Earth and for a smaller duration than for the more massive discs.

Another finding is that for solar-type stars, a secondary gas disc is generated during the luminosity rebound. For a 0.1 \(\text{M}_\oplus\) belt, this secondary gas is almost only composed of photodissociation products. However, such discs might still be detectable (see Figure \ref{fig:flux_jwst_time}). Indeed, the disc luminosity is almost as high as the initial luminosity during this phase. Targeting such a system would be very interesting in order to confirm the secondary nature of the gas. In addition to that, the belt's mass might be higher than assumed here.  We have run another simulation similar to the reference simulation with a belt of one Earth mass (and a viscous parameter \(\alpha = 10^{-3}\)). The gas disc is then more massive and shielded from interstellar photodissociation even several million years after the end of the luminosity rebound.
    
\subsection{Implications for water gas disc observations} \label{sub_seq:water_obs}

\begin{figure}[h]
\begin{centering}
\includegraphics[scale=0.5]{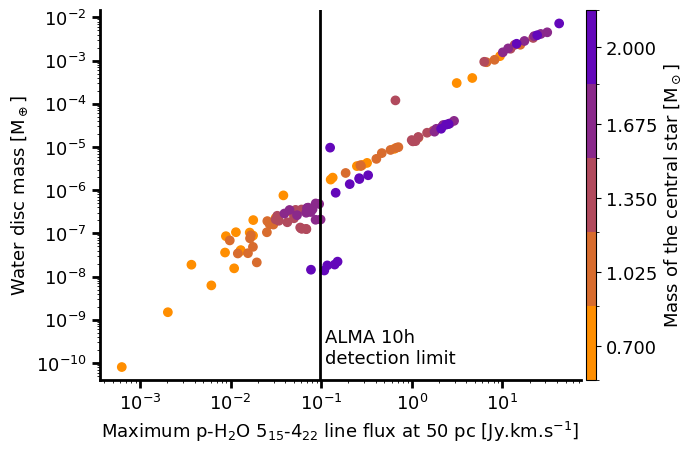}
\caption{\emph{Maximum line flux for the water discs of our simulations at 50 pc for the transition \(\text{pH}_2\text{O} - 5_{15}-4_{22}\) at 325.1529 GHz as a function of the water disc mass and the central star mass. The black vertical line indicates the maximum sensitivity for a 5 \(\sigma\) detection, after 10 hours of observations by ALMA.}}
\label{fig:flux_almap2}

\end{centering}
\end{figure}
    
   The exceptional sensitivity of the ALMA observatory has already enabled the detection of secondary gas discs \citep[e.g.][]{matra_detection_2017,kral_imaging_2019} and, water in protoplanetary discs \citep{facchini_resolved_2024}. The observed system by \cite{facchini_resolved_2024}, HL Tau, is relatively distant (140 pc) and contains an estimated water mass comparable to that found in many of our simulated discs ($\sim$3.6 times the mass of Earth's oceans, i.e.\ $\sim 8 \times 10^{-4}$ Earth masses). To assess which of our discs could be observable with ALMA, we estimated their luminous flux for the transitions $\text{pH}_2\text{O} - 3_{13}-2_{20}$ at 183.31 GHz, $\text{pH}_2\text{O} - 5_{15}-4_{22}$ at 325.1529 GHz, and $\text{oH}_2\text{O} - 10_{29}-9_{36}$ at 321.2257 GHz (see Figures~\ref{fig:flux_almap1}, \ref{fig:flux_almap2}, \ref{fig:flux_almao} for each transition, respectively). For this purpose, we employed the radiative transfer model \texttt{Pythonradex} \citep{cataldi_pythonradex_2026}\footnote{Molecular data \citep{pickett_submillimeter_1998,daniel_collisional_2015} were collected from the EMAA database (\href{https://emaa.osug.fr}{https://emaa.osug.fr}) in February 2026.}. We find that at 50 pc, many discs, including those unshielded from interstellar radiation, are observable within 10 hours of observation. This suggests that massive discs could potentially be observed at much greater distances, thereby increasing the likelihood of detection. Furthermore, it implies that discs can be observed at later stages, particularly if the disc is initially massive and the gas viscosity is low. G-type stars are of particular interest in this context, exhibiting a late production peak at 20-30 Myr.

   Water gas has also already been observed by the JWST \citep[see e.g.,][]{banzatti_jwst_2023}. We also ran radiative transfer simulations using \texttt{pythonradex} \citep{cataldi_pythonradex_2026}\footnote{Molecular data \citep{tennyson_experimental_2001,barber_high-accuracy_2006} were collected in the LAMDA database \citep{schoier_atomic_2005} in February 2026.} for the wavelengths observed by the JWST (see Fig.~\ref{fig:flux_jwst} that includes both NIRSpec and MIRI/MRS transitions, or Figs.~\ref{fig:flux_jwst_nir}, and \ref{fig:flux_jwst_mrs} for NIRSpec and MIRI/MRS, respectively). The JWST sensitivity is superior to that of ALMA for the detection of these discs, and most of the discs studied in this paper are observable at 50 pc. With such high integrated line flux, one can expect to observe discs at several hundred parsecs, even unshielded, low-mass discs. We can also expect to observe older discs with ages close to 100 Myr (see, for example, Figure \ref{fig:flux_jwst_time}, which shows the evolution of the discs' luminosity flux as a function of time for two typical simulations).  With such exceptional sensitivity, one may soon expect a first detection of such a disc with current technology.
In the near future, the ELT-METIS might also be able to detect such discs from the ground, with similar sensitivity, and even ten times better sensitivity between 3 and 4 micron \citep{schmalzl_end--end_2012}, significantly increasing the number of potential systems detectable.

\begin{figure}[h]
\begin{centering}
\includegraphics[scale=0.5]{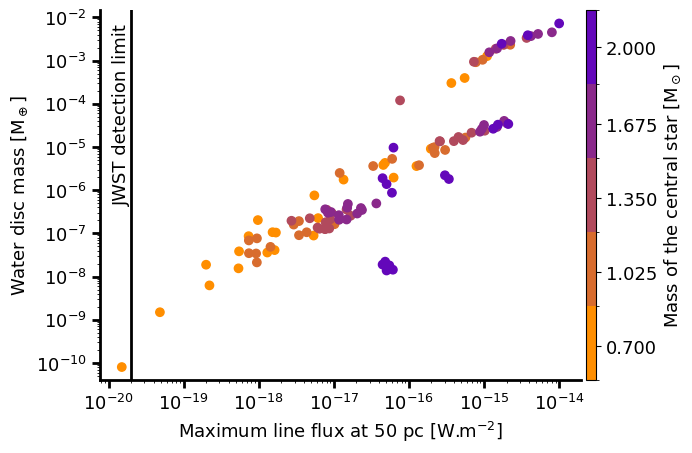}
\caption{\emph{Maximum line flux for the water discs of our simulations at 50 pc in the range observable by JWST/NIRSpec and JWST/MIRI-MRS as a function of the water disc mass and the central star mass. The black vertical line indicates the sensitivity for 10ks ($\sim$ 2h47) to get a 10 \(\sigma\) detection.}}
\label{fig:flux_jwst}
\end{centering}
\end{figure}

 The significance of such observations is major, as they would validate the relevance of this model, particularly in the Solar System, and also help refine it.

\subsection{Accreted water mass onto the inner planet} \label{sub_sec:discuss_accr}

Accretion in our model is highly efficient, particularly due to the inward rediffusion of gas from the outer belt (see Appendix \ref{sub_seq:f_evolution}, Figure \ref{fig:water_accr_mass}). For an Earth-like planet (i.e. with an average \(\text{R}_\text{H}/\text{H}_\text{z}\) ratio of 0.7) in the reference simulation (see Section~\ref{sub_seq:fiducial_result}), we calculate an accreted mass of \(7 \times 10^{-3} \, \text{M}_\oplus\). This is three times greater than the maximum estimated water mass on the Earth \citep[see][]{kral_impact-free_2024}. However, we did not account for the influence of Mars, which would likely reduce this amount by a quarter, or the influence of outer planets, particularly Jupiter, which could decrease it by a factor of 2 (see Section~\ref{sub_seq:res_accr}). This reduces the accreted mass in this case to approximately \(2.6 \times 10^{-3} \, \text{M}_\oplus\), which is close to the current maximum estimated value of \(2.3 \times 10^{-3} \, \text{M}_\oplus\) \citep{kral_impact-free_2024}. Our model is therefore consistent with observations in the Solar System and the conclusions of \cite{kral_impact-free_2024}.

The efficiency of gas production is already high for solar-type stars (G), and we do not expect significant differences in the amount of water accreted by terrestrial planets around more massive stars. The accreted mass will depend partly on the presence of other planets (potentially more massive, such as super-Earths) along the gas path between the belt and the planet, the possible absence of outer planets, and, above all, the mass of the belt. A belt ten times more massive (with a total mass equal to one Earth mass) would result in a tenfold increase in the accreted water mass.

The accretion of this water could significantly alter the surface characteristics of the planet. The extent of this impact will primarily depend on the ability of the planet's rocky mantle to absorb some of this water whether during the cooling phase of a magma ocean or later via plate tectonics. For an Earth mass planet for example, the water mass above which the planet becomes necessarily an ocean planet is around, \(4 \times 10^{-4} \ \text{M}_\oplus\) \citep{guimond_water_2026}. With no others planets, a \(5.2 \times 10^{-3} \, \text{M}_\oplus\) belt would be able to provide such mass. However, the belt mass needed to form ocean planets with other planets in the system might increase by a factor 2 or even more. For a \(3 \, \text{M}_\oplus\) planet, the minimum mass to form an ocean planet would be around \(1.3 \times 10^{-3} \ \text{M}_\oplus\). A belt of \(1.7 \times 10^{-2} \ \text{M}_\oplus\) could then turn a dry planet to an ocean planet.
The absence of plate tectonics or late accretion after the planet's surface has cooled could significantly limit the assimilation of water by the planet's mantle \citep{guimond_water_2026}. In such cases, the amount of water needed to create a planetary ocean is much lower. For Earth, for instance, a mass on the order of one-tenth of the current mass of the oceans (e.g., on the order of \(10^{-5} \, \text{M}_\oplus\)) would suffice to cover the planet, as the topography would also be less pronounced \citep{guimond_water_2026}. Even considering not very optimistic conditions (e.g., a low-mass belt of \(10^{-3} \, \text{M}_\oplus\)), such accreted water masses can be achieved on an inner planet with a mass similar to Earth.

For super-Earths, the \(\text{R}_\text{H} / \text{H}_\text{z}\) ratio is close to 1, allowing even more efficient accretion than for an Earth-like planet (up to 30\% more water by mass). In cases where water is accreted solely at the surface, a water mass on the order of one-tenth of a percent of the planet's mass is sufficient to form an ocean planet \citep[see][]{guimond_blue_2022}. Even with a belt of 0.01 Earth mass (ten times less massive than the most massive belts assumed in our simulations) and outer planets absorbing half of the produced gas mass, an inner planet of 5 Earth masses could become an ocean planet. 

We note that the moment when accretion happens, which is related to the central star's luminosity evolution and to the gas viscosity, is crucial to determine where the water will end up on a rocky planet located in its habitable zone. Too early, most water will be trapped in the planet's core and mantle, and too late, most water accumulates on the surface, thus lowering the threshold to become an ocean planet.

Water also acts as a greenhouse gas, and beyond a critical luminosity threshold from the central star, this greenhouse effect can become runaway \citep[see][]{nakajima_study_1992, turbet_runaway_2019}. For such planets, the radius of their atmospheres increases considerably and is linked to the total amount of water present on the planet \citep{turbet_runaway_2019}. Observing such a planet (via transit) could, when coupled with system formation models, provide a signature of this secondary accretion onto the planet.

\subsection{Uncertainties linked to protostellar formation}

The stellar model we use does not account for the proto-stellar accretion phase. While we focus on exosystems beyond the accretion phase in this study, the long-term stellar evolution may be affected by the modelling of the earliest stages. Our current models are initialised assuming the star has already reached its final mass. Therefore, the star starts as a low-density and cold object, with a high radius and luminosity. Very quickly, the star contracts and the luminosity drops by about a factor of 10, following the well-known Hayashi track. Soon the radiative core develops, and the luminosity slightly increases again with a bump when the ${}^{13}{\rm C}$ fusion ignites, following here the so-called Henyey track.

On the contrary, when considering that the star starts as a seed with a low mass that accretes material until reaching a target mass, the time evolution of the luminosity after the formation can become different, with values always lower than or equal to the classical time evolution \citep{kunitomo_revisiting_2017}. Depending on the scenario assumed for the accretion phase, stringent features like the luminosity peak – which occurs around 20-30~Myrs for a $1~$M$_\odot$ after the proto-stellar accretion phase is over – may be delayed or hastened, enhanced or damped.

Several factors can affect the luminosity path of the model. First, the properties of the initial model are very important. Starting the evolution with a seed of $0.1M_\odot$ \citep{palla_pre-main-sequence_1993} leaves very little imprint on the track after the accretion is over, while the lasting effect (lower luminosity) is more important when starting with a seed of $\simeq 0.001 M_\odot$ \citep[similar to the mass of the second Larson core;][]{kunitomo_revisiting_2017}. Then, two other important ingredients are the accretion rate and its time dependency. With a lower accretion rate, the model reaches the target mass later in the evolution, with a lower luminosity, and the imprint of the accretion will persist longer. \citet{kunitomo_revisiting_2017}, argued that only very small accretion rates (below $10^{-7} M_\odot \cdot {\rm yr}^{-1}$) would produce durable changes in the rest of the pre-main-sequence. For example, for a $1M_\odot$ star, by reducing the accretion rate from $10^{-6}$ to $10^{-7}~M_\odot\cdot{\rm yr}^{-1}$, one can expect a decrease of $\log L/L_\odot$ by $\lesssim 0.1~\rm dex$. The arrival on the main sequence coincides with the end of the bump, without being physically linked to it (as it rather corresponds to the start of the proton-proton chain). The main-sequence would stay unaffected because its dynamic is completely controlled by nuclear reactions. A special case exists, though, for episodic accretion where the star experiences short episodes of vigorous accretion. But in this case, it is hard to derive any conclusion on the time evolution of the luminosity after the accretion phase, as it depends on too many parameters \citep{baraffe_effect_2010,elbakyan_episodic_2019}. Finally, the deuterium abundance of the accreted material can play a role, even after accretion stopped. Deuterium can burn at low temperatures, which can release a lot of energy and create luminosity bursts. Hence, deuterium burning modifies the stellar structure durably and can cause variation to the so-called rebound (or luminosity peak) discussed in Sect. \ref{sub_sec:simulation_set} and visible in Fig. \ref{fig:bol_lum}, at the halfway point of the Henyey track. It coincides with a peak or the end of a plateau in the radius, which then decreases again. In the results presented in \citet{kunitomo_revisiting_2017} for $1 M_\odot$ models, the time at which this second decrease of the radius occurs (i.e. the time of the luminosity peak) is postponed by around 5-10 Myrs by increasing the Deuterium abundance $X_{\rm D}$ from 30 ppm (close to primordial abundance) to 40 ppm. However, the luminosity at the peak is almost unchanged. On the other hand, the peak is almost suppressed in their model when $X_{\rm D} = 0$ ppm. The complete assessment of the impact of the accretion phase on water outgassing shall be left to a dedicated paper.

\section{Conclusions}
    The aim of this study is to generalize the concept of secondary water discs produced in exo-asteroid belts, initially introduced for the Solar System by \cite{kral_impact-free_2024}.
   We have developed a new open-source code \texttt{Diffenix}\footnote{accessible at \url{https://huetpaul.github.io/Diffenix/}} that can model gas release from exo-asteroid belts and accretion onto exoplanets. This new model has new features as it includes a proper treatment of the photodissociation as well as thermal diffusion in the solids.
   We simulated the formation and evolution of such discs across a wide range of configurations, varying parameters such as the mass of the central star, the mass of the asteroid belt, and its location. This enabled us to determine the primary characteristics of these gas discs, particularly with a view to preparing future observations. We are also able to provide an estimate of the mass of water accreted by inner planets during this process. The accreted masses are potentially significant and could have either a positive or negative impact on planetary habitability, depending on the system's configuration.

    \begin{enumerate}
    \item This study confirms the nearly universal nature of the mechanism proposed by \cite{kral_impact-free_2024} for delivering water to the inner planets of the Solar System. Specifically, whenever an asteroid belt is located near the iceline at the time of protoplanetary disc dissipation, a secondary disc forms and water is efficiently accreted onto the inner planets.

    \item The efficiency of gas production is significantly lower around M-type stars, and in these cases, the belt's location is critical in determining the amount of gas produced. Lower water enrichment is expected for planets around such stars. For G-type and more massive stars, production efficiency is very high. There is indeed an important luminosity rebound a few million years after the birth of the central star. This luminosity rebound moves the iceline outward in the asteroid belt (after the dissipation of the primordial disc), increasing water sublimation rate and allowing the sublimation of previously too cold ice in the belt.
    Depending on the system's configuration, the amounts of accreted water can be substantial, potentially enough to form ocean planets.
     
    \item A key aspect of this study concerns the observability of such secondary water gas discs. We find that most discs, including low-mass and unshielded cases, should be detectable with current facilities such as ALMA and the JWST (NIRSpec and MIRI/MRS). With the JWST, water vapour discs may remain observable for up to $\sim$100 Myr after the onset of gas release, and potentially out to distances of several hundred parsecs. As an illustration, the JWST could detect water gas discs with masses as low as $10^{-9}$ M$_\oplus$ at 100 pc, corresponding to the level of gas production expected from belts with initial masses similar to the current Solar System asteroid belt. In the near future, ELT/METIS will further improve these detection limits, further increasing the number of systems accessible to such observations.
    \end{enumerate}
\begin{acknowledgements}
      L. M acknowledge financial support from the French program ‘PROMETHEE’ (Protostellar Magnetism: Heritage vs Evolution) managed by Agence Nationale de la Recherche (ANR-22-CE31-0020).

    This research has made use of spectroscopic and collisional data from the EMAA database 
(\href{https://emaa.osug.fr}{https://emaa.osug.fr} and \href{https://dx.doi.org/10.17178/EMAA}{https://dx.doi.org/10.17178/EMAA}). 
EMAA is supported by the Observatoire des Sciences de l’Univers de Grenoble (OSUG)

    We would also like to thank Martin Turbet and Camille Bergez-Cassalou for the very insightful discussions.
\end{acknowledgements}

\bibliographystyle{aa}
\bibliography{Biblio}
\begin{appendix}\section{Viscous Timescale} \label{sub_seq:viscous_timescale}

Due to their much closer proximity to the central star, the characteristic viscous evolution timescale of water gas discs originating in exo-asteroid belts is expected to differ from those typical of cold secondary CO gas discs. The characteristic viscous diffusion timescale, corresponding to the minimal typical lifetime of the discs, is also likely to vary significantly from one simulation to another.

If $\nu = \alpha c_s \text{H}_\text{z}$ is the kinematic viscosity, with $c_s = \frac{\text{k}_\text{B} T}{\mu  \text{m}_\text{p}}$ the isothermal sound speed and $\text{H}_\text{z} = \frac{c_s}{\Omega_\text{K}} = c_s \sqrt{\frac{a_\text{belt}^3}{\mathcal{G} \text{M}_*}}$ the characteristic scale height of the gas disc at the belt, we can then estimate the characteristic diffusion timescale of the disc :

\begin{equation}
    t_\text{diff} = \frac{\sqrt{G \text{M}_*}}{\alpha} \frac{\mu \text{m}_\text{p} \sqrt{a_\text{belt}}}{\text{k}_\text{B} T}
\end{equation}

We observe that $t_\text{diff}$ evolves inversely proportional to the disc viscosity parameter $\alpha$. Thus, discs with $\alpha = 0.1$ are expected to dissipate a hundred times faster than discs with a viscosity of $\alpha = 10^{-3}$. The dissipation timescale for discs with $\alpha = 10^{-3}$ is on the order of a few million years (see Figure~\ref{fig:test_discs_masses}), compared to a few tens of thousands of years for discs with $\alpha = 0.1$.

\begin{figure}[h]
\begin{centering}
\includegraphics[scale=0.5]{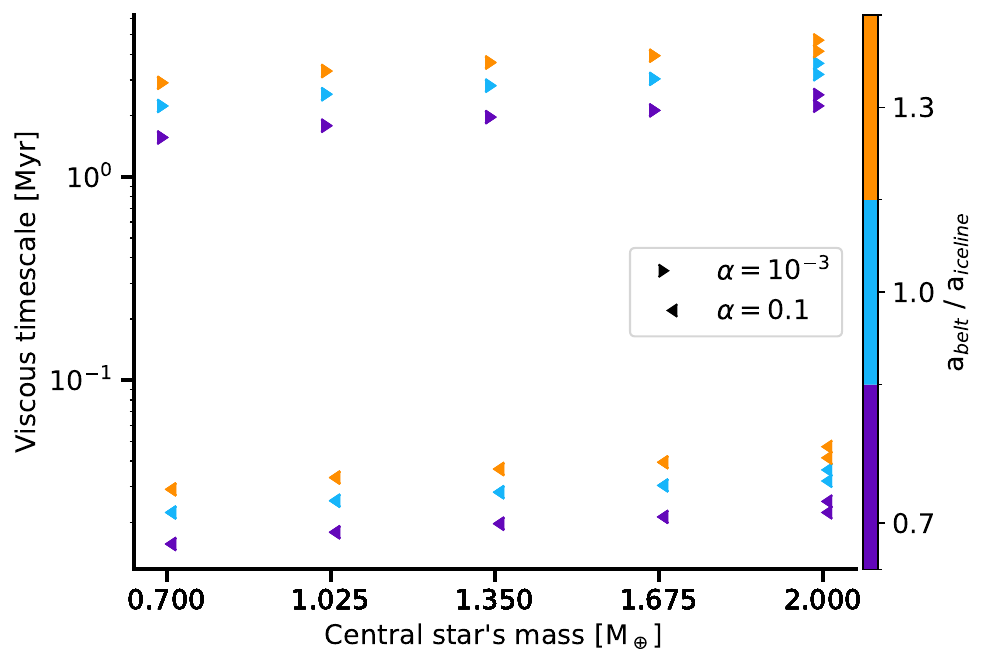}
\caption{\emph{Viscous diffusion timescale for gas discs as a function of the central star mass. The purple, blue, and orange markers represent simulations located 30\% closer, at the iceline, and 30\% further away, respectively. Right-facing triangles represent simulations with low viscosities ($\alpha = 10^{-3}$), while left-facing triangles represent simulations with high viscosities ($\alpha = 0.1$).}}
\label{fig:viscous_timescale}
\end{centering}
\end{figure}

With the radial temperature dependence $\alpha_{T} = -0.5$ chosen in this model, we also find that the viscous dissipation timescale of the discs depends linearly on the semi-major axis of the belt. Thus, discs with their belt located 30\% further from the iceline will dissipate approximately twice as slowly as discs with their belt located 30\% closer to the iceline.

\section{Evolution of the flux primitive $f$} \label{sub_seq:f_evolution}

We initially introduced the variable $f$ to replace the surface density $\Sigma$ for the purpose of simplifying Eq.~\eqref{eq:diff}  such that:
\begin{equation}
    f=\Sigma r^{2+\alpha_T} \label{eq:def_f}
\end{equation}
This choice also proves judicious for simplifying Eq.~\eqref{eq:omega_i}. Moreover, this variable takes on new significance when calculating the mass flux. Indeed, the mass flux at radius $r$ is $\dot{m}(r) = 2 \pi r \, \Sigma v_r$, with $v_r$ the radial velocity defined in Eq.~\eqref{eq:radial_speed}. This yields:
\begin{equation}
    \dot{m}(r) = -6 \pi \sqrt{r} \frac{\partial}{\partial r} \left[ \nu \Sigma \sqrt{r} \right]
\end{equation}

By introducing the viscous parameter $\alpha = \frac{\nu}{c_s \text{H}_\text{z}} = \frac{\mu \text{m}_\text{p} \nu}{\text{k}_\text{B} T} \sqrt{\frac{G \text{M}_*}{r^3}}$, with the temperature $T$ such that $T(r,t) = T(r_0, t) \left(\frac{r}{r_0}\right)^{\alpha_T}$ (see Section~\ref{sub_seq:viscous_diff} for the definition of these quantities) and a new radial coordinate \(x=\sqrt{r}\), we are able to simplify the expression as follows:
\begin{equation}
    \dot{m} = -\frac{3 \pi \nu_0}{x_0^3} \frac{\partial f}{\partial x}
\end{equation}
with $\nu_0 = \nu(r_0) = \frac{\text{k}_\text{B} T(r_0) \, \alpha \, r_0^{3/2}}{\mu \text{m}_\text{p} \sqrt{G \text{M}_*}}$.

We note that if $\alpha_T = -1/2$, as is the case in this study, we can write: $\dot{m} = -4 \pi D \frac{\partial f}{\partial x}$ with $D=\frac{3 \nu_0}{4 r_0^{3/2 + \alpha_T}} x^{1+2\alpha_t}$ the diffusion coefficient. In all cases, $f$ appears as a primitive of the flux.

In this context, it is interesting to consider the form of $f_{st}$ for the steady state of a disc resulting from the generation of gas at a constant rate $\dot{m}_\text{prod}$ at a fixed radius $r_\text{prod}$. This state can be calculated analytically \citep[see, e.g.,][]{kral_imaging_2019}:
\begin{equation}
    f_{st}(x) = \frac{\dot{m}_\text{prod}}{3 \pi \nu(x_\text{prod})} x_\text{prod}^{\alpha_T + 2} \begin{cases}
        \frac{x}{x_\text{prod}} & \text{for } x < x_\text{prod} \\
        1 & \text{for } x > x_\text{prod}
    \end{cases}
\end{equation}
We thus observe that $f_{st}$ is constant in the outer part of the disc, implying that no mass flux is transported outward of the gas production source once steady state is reached. In this situation, all the produced mass is accreted onto the star or onto potential inner planets.

\begin{figure}[h]
\begin{centering}
\includegraphics[scale=0.5]{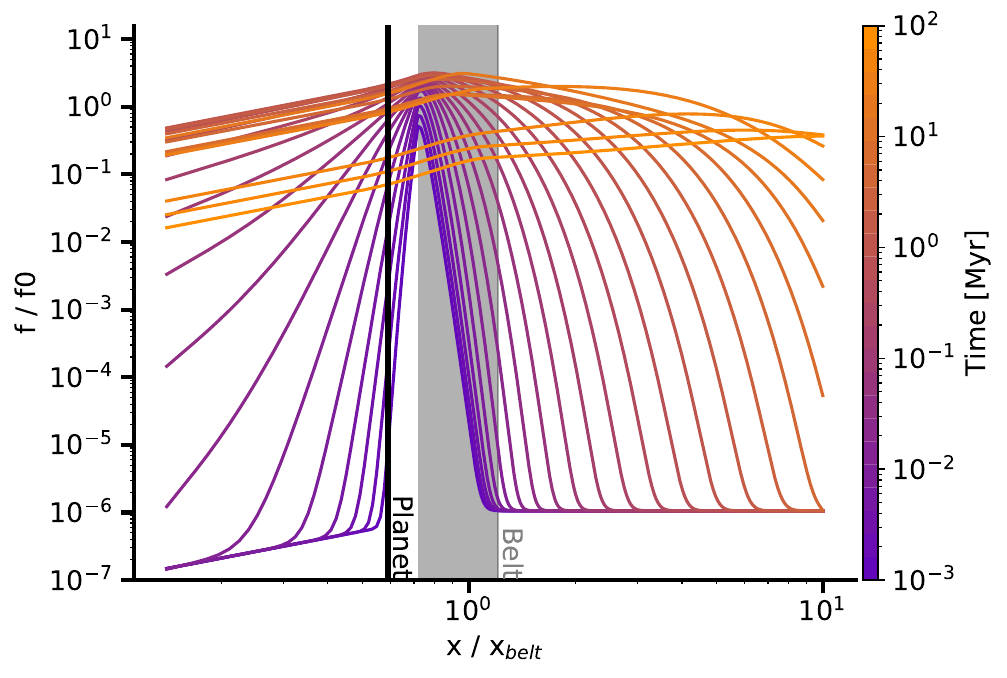}
\caption{\emph{Parameter $f$ as a function of distance from the central star and time (each line colour corresponds to a different time). The abscissa corresponds to the parameter $x$ normalized by $x_\text{belt} = \sqrt{a_\text{belt}}$. The ordinate corresponds to the parameter $f$ arbitrarily normalised by $f_0$. $f_0$ corresponds to the value of $f$ for a steady state with a gas production rate $\dot{m}_{prod} = 10^{-3} \, M_\oplus \cdot \text{Myr}^{-1}$ at the belt. The grey area corresponds to the location of the belt, and the black vertical line to that of the planet.}}
\label{fig:fiducial_f_evolution}
\end{centering}
\end{figure}

The behaviour of $f$ is relatively similar across all simulations. We observe (see Figure~\ref{fig:fiducial_f_evolution}) that it gradually spreads from the belt's position. This is expected since $f$ follows a simple diffusion equation, albeit modulated by accretion onto the inner planet. We can also quickly observe the planet's influence through a drop in $f$ at its location. For the outer part, the initially steep slope of $f$ gradually decreases to become zero, approaching the steady-state case. There is progressively less flux leaving the system as time advances. The major difference with the purely steady-state case is that mass is only produced at the belt for a relatively short duration. Thus, we observe a gradual inversion of the slope of $f$ outside the belt.
This implies that the gas initially diffused outward subsequently diffuses inward. In the context of our simulations, with an infinite integration domain and an infinite integration time, all the produced gas mass should, in principle, accrete onto the inner planet or the star. In practice, if the spatial domain is too small, the gas mass leaving it is definitively lost, and if the integration time is too short compared to the viscous diffusion timescale (see Section~\ref{sub_seq:viscous_timescale}), which is never the case for the simulations in this study, some of the gas mass remains in the disc.

\section{Accretion Cross-Section onto the planets} \label{sub_seq:rh_H}

Due to geometric effects, the accretion by planet of mass $M_p$ orbiting at a distance $a_p$ around a star of mass $\text{M}_*$ and luminosity $L_*$ is limited by the factor $\min(1, \text{R}_\text{H} / \text{H}_\text{z})$ (see Section \ref{sub_sec:planetary_accr}). Here, $\text{R}_\text{H}$ is the Hill radius, and $\text{H}_\text{z}$ the characteristic scale height of the disc (see Section~\ref{sub_sec:planetary_accr}) such that:  
\begin{subequations}
\label{eq:rh_H}
\begin{align}
    \text{R}_\text{H} &= \left(\frac{\text{M}_p}{3 \text{M}_*}\right)^{1/3} \, a_p,  \label{eq:rh}\\
    \text{H}_\text{z} &= \frac{c_s}{\Omega_\text{K}}. \label{eq:hz}
\end{align}
\end{subequations}

The square of the isothermal sound speed in the disc is $c_s^2 = \text{k}_\text{B} T(a_p,\,t)  / (\mu \,\text{m}_\text{p})$, and the Keplerian angular velocity is given by $\Omega_\text{K}^2 = \mathcal{G} \text{M}_*/a_p^3$.

The asteroid belt of this system is located at a distance $a_\text{belt}$ from the central star, and we recall that $a_\text{iceline}$ denotes the position of the iceline at the time when the protoplanetary disc dissipates.

We can then express the ratio $\text{R}_\text{H} / \text{H}_\text{z}$ in terms of the characteristic quantities that follow :

\begin{equation}
    \begin{split}
         \frac{\text{R}_\text{H}}{\text{H}_\text{z}}(r_p, t) &= S_0 \left(\frac{a_\text{iceline}}{a_\oplus}\right)^{-\frac{1 + \alpha_T}{2}} \left(\frac{\text{L}_*}{\text{L}_\odot}\right)^{-1/8}  \left(\frac{\text{M}_p}{\text{M}_\oplus}\right)^{1/3} \\
         & \left(\frac{\text{M}_*}{\text{M}_\odot}\right)^{1/2} \left(\frac{a_\text{belt}}{a_\text{iceline}}\right)^{-\frac{1 + \alpha_T}{2}}  \left(\frac{r_p}{a_\text{belt}}\right)^{-\frac{1 + \alpha_T}{2}}
    \end{split}
\end{equation}

\noindent with $S_0 = M_\oplus^{1/3} \sqrt{\frac{\mathcal{G} \,\mu \text{m}_\text{p} \, \text{M}_\odot}{3^{2/3} \, \text{k}_\text{B} \, \text{T}_\oplus \ a_\oplus}} \simeq 0.53$ and $\text{T}_\oplus = 278 \, \text{K}$ (i.e. assuming a black body temperature for the gas), $a_\oplus = 1 \, \text{au}$.

In the case of our simulations, the factor $\frac{r_p}{a_\text{belt}}$ is fixed at 0.8. Thus, for all our simulations, $\left(\frac{r_p}{a_\text{belt}}\right)^{-\frac{1 + \alpha_T}{2}} \simeq 1.06$. The factor $\frac{a_\text{belt}}{a_\text{iceline}}$ is either 0.7, 1, or 1.3. This implies that for our simulations, $\left(\frac{a_\text{belt}}{a_\text{iceline}}\right)^{-\frac{1 + \alpha_T}{2}}$ is approximately 1.09, 1, and 0.93, respectively.

The position of the iceline depends non-trivially on the mass of the central star, while the luminosity of the star also varies over time. To estimate the ratio $\text{R}_\text{H} / \text{H}_\text{z}$ in our simulations, we can refer to Figure~\ref{fig:rh_h}.

\begin{figure}[h]
\begin{centering}
\includegraphics[scale=0.5]{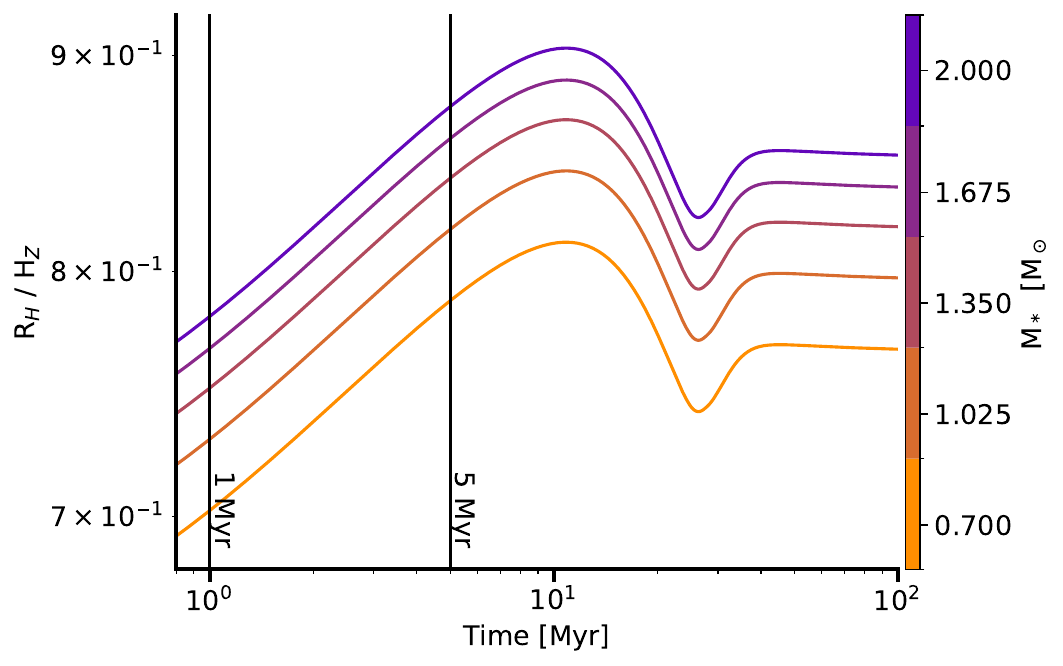}
\caption{\emph{Ratio  $\text{R}_\text{H} / \text{H}_\text{z}$ as a function of time and central star mass for an Earth-mass planet in a system with a belt located at the iceline. The origin of time is the origin of the system}. The two black vertical bars indicate the initial times chosen for our simulations.}
\label{fig:rh_h}
\end{centering}
\end{figure}

\section{Estimation of the UV fluxes in stellar models}
\label{section:xuv}

The XUV fluxes can be estimated following empirical relations in a saturated and a unsaturated dynamo regimes, provided by \citet{wright_stellar-activity-rotation_2011} (hereafter W11) or \citet{johnstone_active_2021} (hereafter J21), working on the same data. Both groups proposed a relation between $R_\xX = L_\xX / L_{\rm bol}$, the ratio of X-to-bolometric luminosities and the Rossby number $\Ross = P_{\rm rot} / \tau_{\rm conv}$, with $\tau_{\rm conv}$ the convective turnover time. It is worth noting, that this last quantity is derived by W11 or J21 with a scaling relation calibrated on grids of stellar models different, while it is directly accessible within \cesamxx using MLT. The fitting functions proposed by W11 is
\begin{equation}
    R_{\xX, W11} = \left\{
    \begin{array}{lr}
        C \Ross^\beta    & \textrm{if} \Ross > \Ross_{\rm sat} \\
        R_{\xX, \rm sat} & \textrm{if} \Ross \leq \Ross_{\rm sat} \\
    \end{array}\right.,
\end{equation}
with $\log R_{\rm X, sat} = -3.13 \pm 0.08$, $\Ross_{\rm sat} = 0.13\pm 0.02$, $\beta = -2.70\pm 0.13$ and $C = R_{\xX, \rm sat} / \Ross_{\rm sat}^\beta$. J21 proposed a slightly different scaling relation as
\begin{equation}
    R_{\xX, J21} = \left\{
    \begin{array}{lr}
        C_1 \Ross^{\beta_1}    & \textrm{if} \Ross > \Ross_{\rm sat} \\
        C_2 \Ross^{\beta_2}    & \textrm{if} \Ross \leq \Ross_{\rm sat} \\
    \end{array}\right..
\end{equation}
with $\Ross_{\rm sat} = 0.0605\pm 0.0031$, $\beta_1 = -0.135\pm0.030$, $\beta_2 = -1.889\pm0.079$, $C_1 = R_{\xX, \rm sat} / \Ross_{\rm sat}^{\beta_1}$, $C_2 = R_{\xX, \rm sat} / \Ross_{\rm sat}^{\beta_2}$ and $R_{\xX, \rm sat} = 5.135\times10^{-4} \pm 3.320 \times 10^{-5}$. In addition, J21 derive similar scaling relations for the UV fluxes in the 10-36 nm (EUV1) and  36-92 nm (EUV2) bands. One can simply express the surface flux $F_{\rm EUV1,2} = L_{\rm EUV1,2} / (4\pi R_*^2)$ in each band with 
\begin{equation}
    \log F_{\rm EUV1} = 2.04 + 0.681 \log F_{\rm X}
\end{equation}
and
\begin{equation}
    \log F_{\rm EUV2} = -0.341 + 0.920 \log F_{\rm EUV1},
\end{equation}
where $R_*$ is the stellar radius and $F_{\rm X} = L_{\rm bol} R_{\rm X} / (4\pi R_*^2)$.

\section{Extra Figure}

\begin{figure}[h]
\begin{centering}
\includegraphics[scale=0.5]{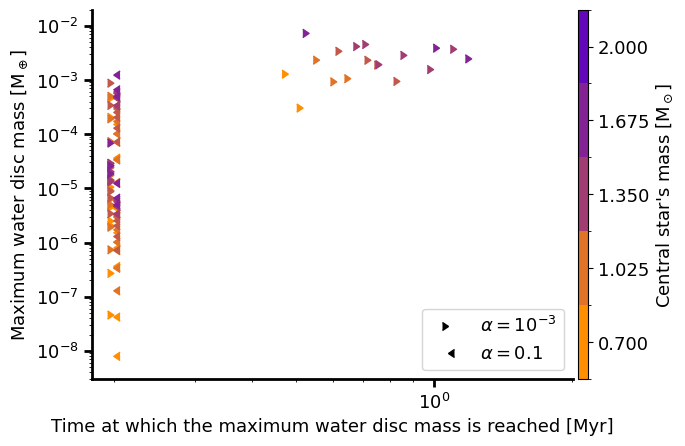}
\caption{\emph{Maximum mass of water vapour discs as a function of the time at which this mass is reached after $t_{\text{init}}$. The colours are associated with the mass of the central star, while the right and left arrows indicate that the gas viscosity is $10^{-3}$ and 0.1, respectively. Only simulations with a massive belt (mass $M_\text{belt} = 0.1 \, M_\oplus$)} are represented.}
\label{fig:diag_mwater_twater}
\end{centering}
\end{figure}

\begin{figure}[h]
\begin{centering}
\includegraphics[scale=0.5]{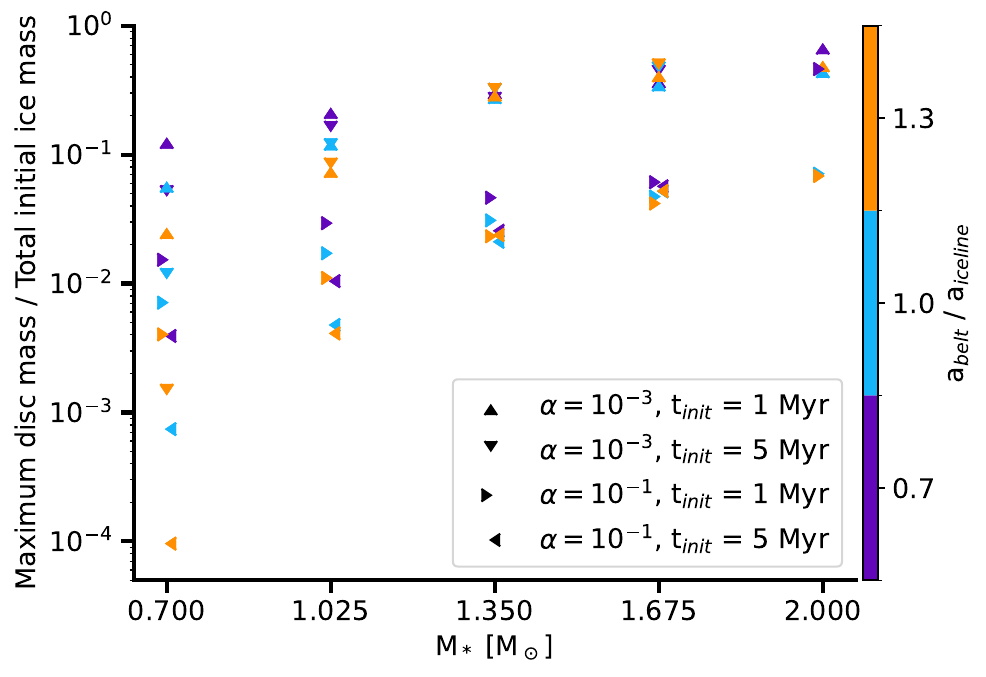}
\caption{\emph{Maximum disc mass compared to the total initial ice mass as a function of the central star's mass. The purple, blue and orange points are simulations at 0.7, 1.0, and 1.3 times the ice-line, respectively. The right and left triangles represent simulations with a viscous parameter \(\alpha = 0.1\) and an initial time of 1~Myr and 5~Myr, respectively. The upward and downward triangles represent simulations with a viscous parameter \(\alpha = 10^{-3}\) with initial times of 1~Myr and 5~Myr, respectively.}}
\label{fig:diag_mmax}
\end{centering}
\end{figure}

\begin{figure}[h]
\begin{centering}
\includegraphics[scale=0.5]{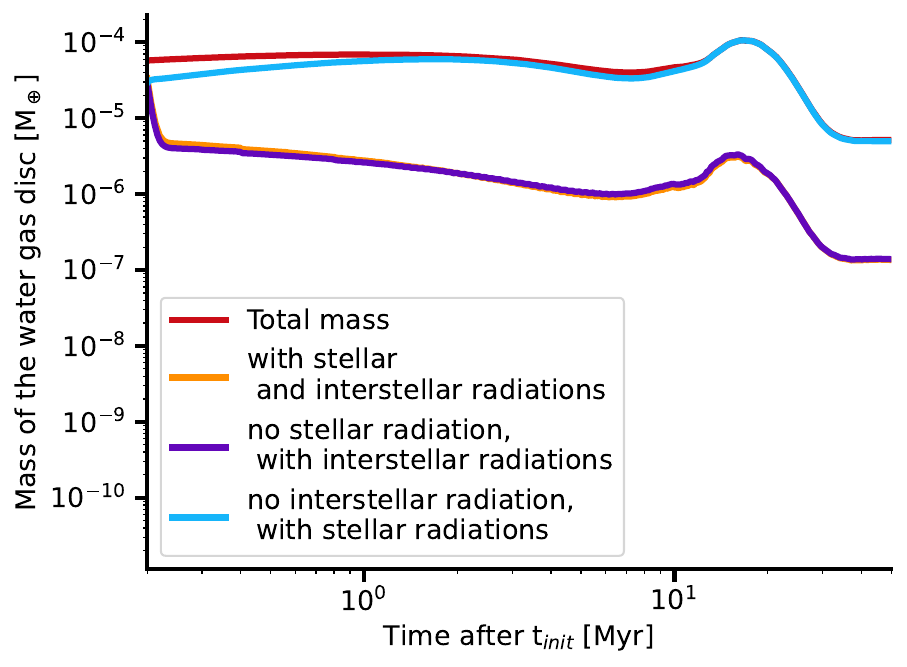}
\caption{\emph{Total mass and water vapour mass of the disc in the reference simulation, shown in red and orange respectively. The purple line represents the water disc mass for the simulation without stellar radiation, and the blue line represents the water disc mass for the simulation without interstellar radiation.}}
\label{fig:test_discs_masses}
\end{centering}
\end{figure}

\begin{figure}[h]
\begin{centering}
\includegraphics[scale=0.5]{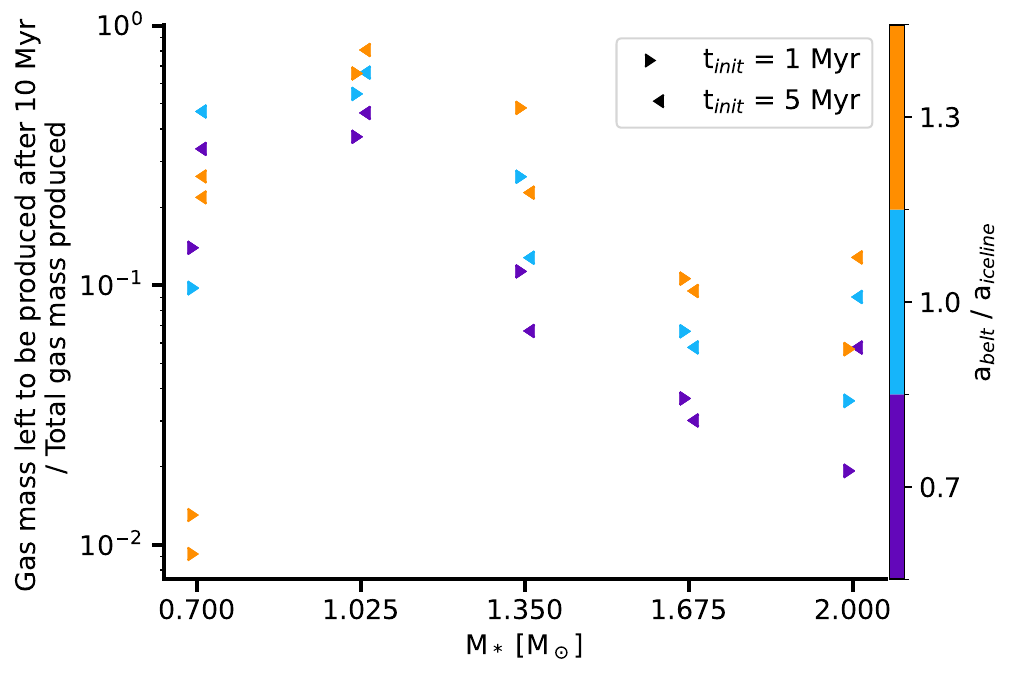}
\caption{\emph{Fraction of gas remaining to be produced relative to the total gas mass produced, 10~Myr after the start of the simulation \(t_{\text{init}}\), as a function of the central star's mass. The colours associated with the colourbar denote the distance of the belts from the central star relative to the ice-line.}}
\label{fig:diag_mprod_time}
\end{centering}
\end{figure}

\begin{figure}[h]
\begin{centering}
\includegraphics[scale=0.5]{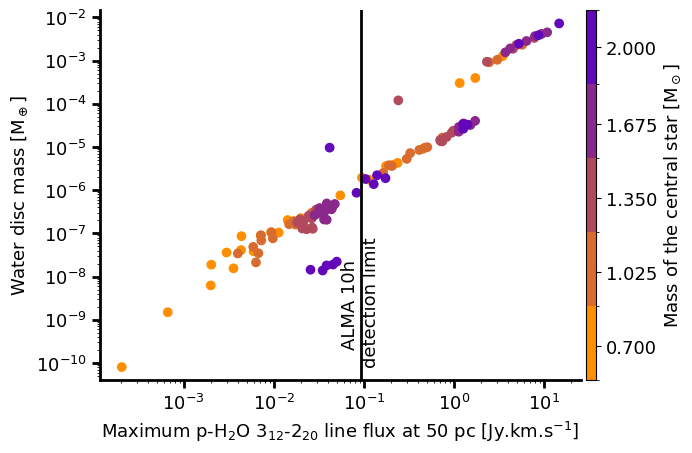}
\caption{\emph{Maximum line flux for the water discs of our simulations at 50 pc for the transition \(\text{pH}_2\text{O} - 3_{13}-2_{20}\) at 183.31 GHz as a function of the water disc mass and the central star mass. The black vertical line indicates the maximum sensitivity for a 5 \(\sigma\) detection, after 10 hours of observations by ALMA.}}
\label{fig:flux_almap1}
\end{centering}
\end{figure}

\begin{figure}[h]
\begin{centering}
\includegraphics[scale=0.5]{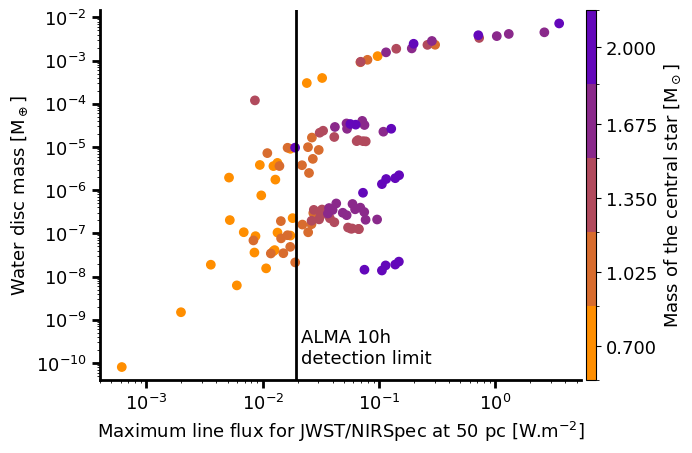}
\caption{\emph{Maximum line flux for the water discs of our simulations at 50 pc for the transition \(\text{oH}_2\text{O} - 10_{29}-9_{36}\) at 321.2257 GHz as a function of the water disc mass and the central star mass. The black vertical line indicates the maximum sensitivity for a 5 \(\sigma\) detection, after 10 hours of observations by ALMA.}}
\label{fig:flux_almao}
\end{centering}
\end{figure}

\begin{figure}[h]
\begin{centering}
\includegraphics[scale=0.5]{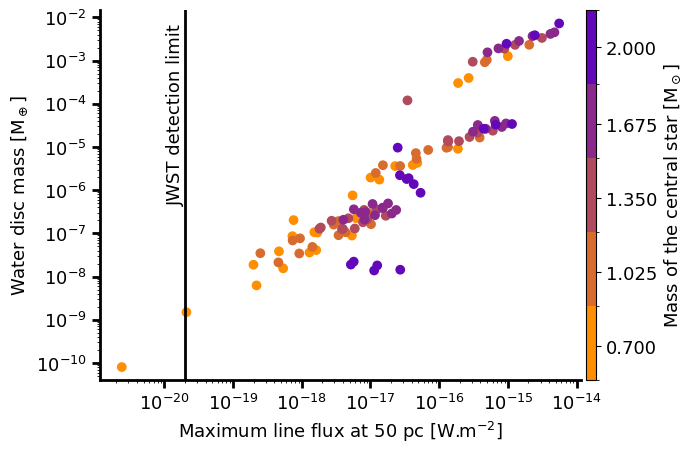}
\caption{\emph{Maximum line flux for the water discs of our simulations at 50 pc in the range observable by JWST/NIRSpec as a function of the water disc mass and the central star mass. The black vertical line indicates the sensitivity for 10ks ($\sim$ 2h47) to get a 10 \(\sigma\) detection. }}
\label{fig:flux_jwst_nir}
\end{centering}
\end{figure}

\begin{figure}[h]
\begin{centering}
\includegraphics[scale=0.5]{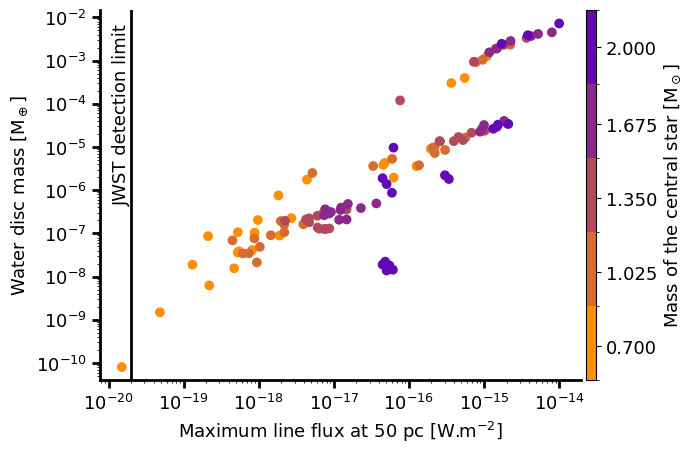}
\caption{\emph{Maximum line flux for the water discs of our simulations at 50 pc in the range observable by JWST/MIRI-MRS as a function of the water disc mass and the central star mass. The black vertical line indicates the sensitivity for 10ks ($\sim$ 2h47) to get a 10 \(\sigma\) detection. }}
\label{fig:flux_jwst_mrs}
\end{centering}
\end{figure}

\begin{figure}[h]
\begin{centering}
\includegraphics[scale=0.5]{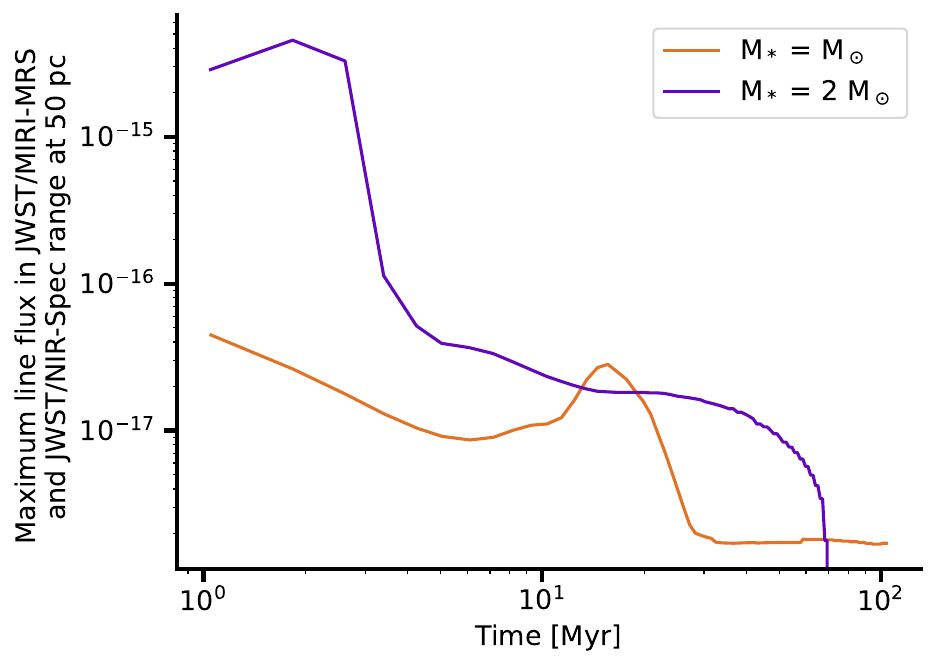}
\caption{\emph{The maximum line flux for the water discs at 50 pc, as observed by JWST/MIRI-MRS and JWST/NIRSpec, is shown as a function of time and central star mass (purple: 1 solar mass; orange: 2 solar masses). The two simulations represented here have a 0.1 Earth mass belt located at the ice line and start 5 million years after the initial formation of the star.}}

\label{fig:flux_jwst_time}
\end{centering}
\end{figure}

\end{appendix}
\end{document}